\documentclass[12pt]{article}
\usepackage{a4wide}
\usepackage{amsmath}
\newcommand{\meno}{\textrm{--}\,}
\newcommand{\out}{\textrm{\scriptsize{out}}}
\newcommand{\inn}{\textrm{\scriptsize{in}}}
\newcommand{\disc}{\textrm{\scriptsize{disc}}}
\newcommand{\conn}{\textrm{\scriptsize{conn}}}

\makeatletter

\newcommand{\rj}{\Re\langle J \rangle}

\newcommand{\beq}{\begin{equation}}
\newcommand{\eeq}{\end{equation}}
\newcommand{\bea}{\begin{eqnarray}}
\newcommand{\eea}{\end{eqnarray}}
\newcommand{\brr}{\begin{array}}
\newcommand{\err}{\end{array}}
\newcommand{\bc}{\begin{center}}
\newcommand{\ec}{\end{center}}

\let\SF@@footnote\footnote \def\footnote{\ifx\protect\@typeset@protect
\expandafter\SF@@footnote \else \expandafter\SF@gobble@opt \fi }
\expandafter\def\csname SF@gobble@opt \endcsname{\@ifnextchar[
  \SF@gobble@twobracket \@gobble }
  \edef\SF@gobble@opt{\noexpand\protect \expandafter\noexpand\csname
  SF@gobble@opt \endcsname} \def\SF@gobble@twobracket[#1]#2{}

\def\Journal#1#2#3#4{{#1} {\bf #2} (#4) #3}

\def\NPB{\em Nucl. Phys.}
\def\PLB{\em Phys. Lett.}

\def\CMP{\em Commun. Math. Phys.}

\newcommand{\pdir}{p\kern -5.2pt\raise 0.2ex\hbox {/}}
\newcommand{\vdir}{v\kern -5.75pt\raise 0.15ex\hbox {/}}
\newcommand{\kdir}{k\kern -5.75pt\raise 0.15ex\hbox {/}}
\newcommand{\epsdir}{\epsilon\kern -5.0pt\raise 0.15ex\hbox {/}}
\newcommand{\bvdir}{\bar{v}\kern -5.75pt\raise 0.15ex\hbox {/}}
\newcommand{\Ddir}{D\kern -7.75pt\raise 0.20ex\hbox {/}}
\newcommand{\ldir}{l\kern -5.0pt\raise 0.2ex\hbox{/}}
\newcommand{\varepsdir}{\varepsilon\kern -5.5pt\raise 0.15ex\hbox{/}}

\newcommand{\state} {|\,\pi(\vec p\,)\,\pi(-\vec p\,)\,\rangle_{\textrm{in}}}
\makeatother

\begin{document}
\begin{flushright}
 {ROMA-1395/00} \\
{SHEP 01/10}\\
hep-lat/0104006
\end{flushright}
\vskip 1.5cm
\begin{center}
\par \textbf {\Large
\boldmath{$K \to \pi\pi$} Decays in a Finite Volume}

\vskip 0.7cm
\textbf{C.-J.D.~Lin$^a$, G.~Martinelli$^b$, C.T.~Sachrajda$^a$
and M.~Testa$^b$}

\vskip 0.2cm

$^a$\textit{Dept. of Physics and Astronomy, Univ. of Southampton,\\
Southampton, SO17 1BJ, UK}

$^b$\textit{Dip. di Fisica, Univ. di Roma ``La Sapienza'' and INFN,
Sezione di Roma,\\
Piazzale Aldo Moro 2, I-00185 Rome, Italy}
\end{center}

\vskip 0.75cm

\begin{abstract}We discuss finite-volume computations of two-body
hadronic decays below the inelastic threshold (e.g.  $K\to\pi\pi$
decays).  The relation between finite-volume matrix elements and
physical amplitudes, recently derived by Lellouch and L\"uscher, is
extended to all elastic states under the inelastic threshold.  We
present a detailed comparison of our approach with that of Lellouch
and L\"uscher and discuss the possible limitations of the method which
could arise due to the presence of inelastic thresholds. We also
examine a standard alternative method which can be used to extract the real
part of the decay amplitude from correlators of the form
$\langle\,0\,|T[\pi\pi{\cal H}_WK\,]|\,0\,\rangle$. We show that in
this case there are finite-volume corrections which vanish as inverse
powers of the volume, which cannot be removed by a multiplicative
factor.
\end{abstract}

\vfill\today
\newpage \clearpage \setcounter{page}{1}

\section{Introduction} \label{sec:introduction}

The theory of non-leptonic weak decays represents a major challenge
for particle physicists because of our inability to control the strong
interaction effects.  Lattice QCD offers a natural opportunity to
compute the corresponding matrix elements from first principles.  A
number of theoretical questions however, needs to be answered before
such computations can be performed with sufficient precision to be
phenomenologically relevant. The main difficulties are related to the
continuum limit of the regularized theory (the \textit{ultra-violet}
problem) and to the relation between matrix elements computed in a
finite Euclidean space-time volume and the corresponding physical
amplitudes (the \textit{infrared} problem). The ultra-violet problem,
which deals with the construction of finite matrix elements of
renormalized operators constructed from the bare lattice ones, has
been addressed in a series of papers~\cite{renorm}-\cite{dawson} and
we will not consider it further in this work. The infrared problem
arises from two sources:
\begin{itemize}
\vspace{-0.2in}\item the unavoidable continuation of the theory to
Euclidean space-time and
\item the  use of a finite volume in numerical simulations.
\end{itemize}
\vspace{-0.2in}An important step towards the solution of the infrared
problem has recently been achieved by Lellouch and L\"uscher~\cite{lll} 
(denoted in the following by LL), who derived a
relation between the $K\to\pi\pi$ matrix elements in a finite volume
and the physical kaon-decay amplitudes. This relation is valid up to
exponentially vanishing finite-volume effects.

In this paper we present an alternative discussion of boundary effects
and the LL-formula, based on a study of the properties of correlators
of local s-wave operators~\footnote{s-wave operators are those which,
when applied to the vacuum in a finite volume, excite only states with
an s-wave component. They include the finite-volume restriction of all
infinite-volume rotationally invariant local operators.} We derive
LL-formulae for all elastic states under the inelastic threshold, with
exponential accuracy in the quantization volume. A consequence of our
derivation is that the relation between finite-volume matrix elements
and physical amplitudes, derived by Lellouch and L\"uscher for the
lowest seven energy eigenstates can be extended to all elastic states
under the inelastic threshold. Our analysis also demonstrates
explicitly how finite volume correlators converge to the corresponding
ones in infinite volume.

Our approach is based on the property of correlators of local
operators which can be expressed, with exponential accuracy, both as a
sum or as an integral over intermediate states, when considering
volumes larger than the interaction radius and Euclidean times $0<t
\simeq L$. A possible source of concern is therefore the (practically
relevant) situation in which the quantization volume allows only
very few (perhaps two or three) elastic states under the inelastic
threshold. In this case we will show that it is not possible, in
general, to derive the LL-formula from the behaviour of correlators of
local operators. At first sight this may appear to be a weakness of
our approach compared to that of ref.~\cite{lll}.  We will argue
however, that the validity of the L\"uscher quantization condition for
the energy levels of two-particle states in a finite volume~\cite{ml}
also requires the volume to be sufficiently large for the Fourier
series to be equal to the infinite-volume energy integrals, up to
exponential corrections. Since the derivation of the LL relation in
ref.~\cite{lll} relies on this quantization formula, we conclude that
the conditions on the volume for the applicability of this relation
are equivalent in the two approaches. This issue is of considerable
practical significance. For the foreseeable future, in actual lattice
simulations of $K\to\pi\pi$ decays there will be very few elastic
states below the inelastic threshold, $E_{th}$ (perhaps even only two
such states).  It is therefore necessary to examine the precision with
which the integral over the energy is approximated by the sum over the
elastic finite-volume energy levels below $E_{th}$.  In this paper we
discuss this question theoretically.

In spite of the limitations discussed in the previous paragraph, we
will argue that for an important class of dynamical situations the
proximity of the kinematic inelastic threshold does not necessarily
invalidate the LL-relation and that these situations are likely to
include the two-pion system in physical kaon decays.  A more
quantitative answer will have to await results from detailed numerical
simulations in the future. The number of states below $E_{th}$ can be
changed by varying either the masses of the mesons in a given volume
(and hence considering unphysical decays) or the size of the
volume. Thus one has the opportunity of investigating the importance
of finite-volume corrections in exclusive two-body decays
quantitatively.

Another important consequence of our approach is the demonstration
that it is possible to extract $K \to \pi\pi$ amplitudes also when the
kaon mass, $m_K$, is not equal to the two-pion energy, i.e. when the
inserted weak Hamiltonian operator carries a non-zero
energy-momentum. Such amplitudes may be useful, for example, in the
determination of the coefficients of the operators appearing in the
chiral expansion.

We also examine an alternative method often used to obtain information
about $K\to\pi\pi$ decay amplitudes. This method is based on the
evaluation of the correlation function\\ $\langle\,0\,|\pi_{\vec
q}(t_1)\pi_{-\vec q}(t_2){\cal H}_{W}(0) K(t_K)\,|\,0\, \rangle$, 
where $\pi_{\vec q}(t)$ denotes
the three dimensional Fourier transform of the pion's interpolating
field $\pi(x)$~\footnote{In the introduction we denote by $K$ and
$\pi$ interpolating fields with the quantum numbers of these
particles. In subsequent sections we reserve $K$ and $\pi$ to denote
the corresponding particles and introduce a different notation for the
interpolating fields.}. Taking the times $t_{1,2}$ to be large and
positive with $t_1\gg t_2$ and $t_K$ to be large and negative, we
obtain the real part of the physical $K\to\pi\pi$ decay amplitude (see
below), up to power corrections in the volume. These power corrections
cannot be removed by multiplication by the LL-factor.

The remainder of this paper is organised as follows. In
sec.~\ref{sec:mt} we start by recalling some basic facts about
infinite-volume Euclidean Green functions related to kaon decays. In
particular, we stress the result derived in ref.~\cite{mt} that in
Euclidean space one obtains the average of the matrix elements into
\textit{in} and \textit{out} two-pion states. We expand the discussion
of the similarities and differences between the Euclidean- and
Minkowski-space calculations in appendix A, where we demonstrate how
the physical $K\to\pi\pi$ amplitude (for which the two-pion state is
an
\textit{out} state) is recovered in Minkowski space.

We then proceed to discuss calculations in a finite volume.  In
section \ref{3D} we present a heuristic discussion concerning
correlators in  a three dimensional periodic cubic volume.  In
sec.~\ref{sec:large} the finite volume formulae are derived and a
detailed comparison of our approach with that of ref.~\cite{lll}
is presented. In this section we also examine the conditions on
the volume required for our derivation and in sec.~\ref{sec:wave}
we analyze those needed for the L\"uscher quantization
formula~\cite{ml}.

In sec.~\ref{sec:delta} we present an alternative method to obtain
information about the kaon decay amplitude, based on the correlator
$\langle\,0\,|\pi_{\vec q}(t_1)\pi_{-\vec q}(t_2){\cal H}_{W}(0)
K(t_K)\,|\,0\, \rangle$ introduced above. We show that in this case
there are power corrections in the volume and explain why they are not
removable by the LL correcting factor. We present our conclusions in
section~\ref{sec:conclusion}.

There are three further appendices, introduced to clarify the
discussion. In appendix B we demonstrate that it is indeed $\Re\{{\cal
A}\}$ which is obtained from the $\langle\,0\,|T[\pi\pi{\cal
H}_{W}K]\,|\,0\,\rangle$ correlator ($\Re\{{\cal A}\}$ represents the
real part of $A$). In Appendix C we sketch the derivation of the
L\"uscher quantization formula and some basic facts needed in the
text. In appendix D we demonstrate that  it is possible to determine
the scattering phase $\delta$ directly by computing four-pion
correlation functions.

\section{Euclidean Green Functions and Physical Amplit\-udes}
\label{sec:mt}

One of the main obstacles in the extraction of physical amplitudes
from lattice simulations stems from the rescattering of final
state particles in Euclidean space. The formalization of this
problem, in the infinite-volume case, was considered in
ref.~\cite{mt} and is referred to as the {\it Maiani-Testa no-go
theorem}. In this section we review the arguments leading to this
theorem, elucidating the main differences between the Euclidean
and Minkowski formulations of the theory. This allows us to fix
the notation and to introduce several quantities which will be
used in the following. The discussion of finite-volume effects is
postponed to the following sections.

We are ultimately interested in $K\to \pi \pi$ decays.  Single
particle states however, do not present any theoretical difficulty
so, following ref.~\cite{mt}, in order to simplify the discussion
while retaining the essential physical aspects, we eliminate the
kaon and consider the Green function
\begin{equation}
{\cal G}(t_1,t_2;\vec q)=\langle 0\,|\,\Phi_{\vec
q\,}(t_1)\,\Phi_{\text{-}\vec q\,}(t_2) \,J(0)\,|\,0\rangle\ ,
\label{eq:gdef}\end{equation} where $t_1>t_2>0$; $J$ is a local
operator which can create two-pion states with a definite isospin,
from the vacuum;
\begin{equation}
\Phi_{\vec q\,}(t)\equiv\int\,d^{\,3}\!x\,e^{-i\vec q\cdot\vec
x}\, \phi(\vec x,t) \, , \label{eq:Phidef}\end{equation} and
$\phi(\vec x,t)$ is some appropriately chosen interpolating
operator for the pion. For convenience the two-pion intermediate
state is chosen to have total momentum equal to zero. From the
continuation of the correlation function (\ref{eq:gdef}) to
Minkowski space, using LSZ reduction formulae one can determine
the form factor
\begin{equation}
_{\textrm{\scriptsize{out}}}\langle\pi(\vec q)\,\pi(\textrm{--}\,\vec q)\,
|\,J(0)\,|0\rangle\ ,
\label{eq:ff}\end{equation}
which we assume to be the goal of some numerical simulation. In order
to avoid unnecessary kinematical complications, in the following we
discuss the case of a Hermitian, local, scalar operator $J$, which
excites zero angular-momentum states from the vacuum. The extension of
the present discussion to correlators relevant for $K \to \pi \pi$
decays is completely straightforward.

In practice, in a numerical simulation the Euclidean correlation
functions can be computed only approximately and so the continuation
to Minkowski space is impossible.  We therefore have the problem of
extracting the relevant physical information from the Euclidean
correlation functions.

The general expression of eq.~(\ref{eq:gdef}) in terms of matrix
elements was derived in ref.~\cite{mt}. For $t_1\gg t_2 > 0$ one has
\begin{eqnarray}
{\cal G}(t_1,t_2;\vec q\,) & = &
\frac{Z}{(2E_q)^2}\,e^{-E_q(t_1+t_2)} \bigg\{\, \frac{1}{2} \Big[\
_{\textrm{\scriptsize{out\!}}}\langle\pi(\vec q\,)\pi(\meno\vec
q\,) \,|\,J(0)\,|0\rangle\, +\,
_{\textrm{\scriptsize{in}}}\langle\pi(\vec q\,)\pi(\meno\vec q\,)
\,|\,J(0)\,|0\rangle\, \Big] \nonumber\\ && \hspace{2in}+\
2E_q{\cal P}_{\vec q\,}(t_2)\, \bigg\}\ ,
\label{eq:mtresult}\end{eqnarray} where \bea {\cal P}_{\vec
q\,}(t_2)&\equiv& {\cal P}\,\sum_n\, \langle\,\pi(\vec
q\,)\,|\,\Phi_{-\vec q\,}(0)\,|n\,\rangle_{\textrm{out}}
^{\textrm{conn}}\ _{\textrm{out}}\langle\,n\,|\,J(0)\,|0\rangle\,
e^{-(E_n-  2\,E_q)t_2}\, , \label{eq:pp}\\ E_q& =&\sqrt{\vec q^{\
2} + m_\pi^2} \quad \quad \textrm{and}\quad\quad \sqrt{Z}=\langle
0|\,\phi(\vec 0,0)|\pi(\vec q\,)\rangle\, . \label{eq:ezdefs}\eea
$\sqrt Z$ and $E_q$ can be obtained by computing the propagator of
a single pion. We use the normalization $\langle\pi(\vec q\,)
\vert \pi(\vec p\,)\rangle = (2 \pi)^3 2 E_q \delta^3(\vec q -
\vec p\,)$.  ${\cal P}$ represents the principal value of the
integral  implicit in the $\sum_{n}$ and the superscript {\small
``conn''} implies that the corresponding matrix element is the
connected one. Although the term in square brackets in
eq.~(\ref{eq:mtresult}) is real and hence cannot be equal to the
form factor in eq.~(\ref{eq:ff}), it is nevertheless a physical
quantity which can be directly compared with experiment.  To see
this it is convenient to write \bea \Re\langle\, J\, \rangle
&\equiv& \frac{1}{2} \Big[\
_{\textrm{\scriptsize{out\!}}}\langle\,\pi(\vec q\,)\pi(\meno\vec
q\,) \,|\,J(0)\,|\,0\,\rangle\, +\,
_{\textrm{\scriptsize{in}}}\langle\pi(\vec q\,)\pi(\meno\vec q\,)
\,|\,J(0)\,|0\,\rangle\, \Big] \nonumber \\ &=&
\rule[-0.1cm]{0cm}{0.6cm}\hspace{0.2cm}
\big|\,_{\textrm{out}}\langle\,\pi(\vec q\,) \pi(-\vec
q\,)\,|\,J(0)\,|0\,\rangle\,\big|\,\cos\delta(2 E_{q}) \ ,
\label{eq:pp1} \eea where $\delta(2 E_{q})$ is the FSI phase shift
for the two pions. Here and in the following we assume that the
two-pion energy is below the inelastic threshold. Unfortunately,
the extraction of $\Re\langle\, J\,\rangle$ from the correlation
function at large time distances is impossible in practice, since
in  this limit it is exponentially dominated by the last term in
eq.~(\ref{eq:mtresult}). The conclusion of ref.~\cite{mt} was
therefore that it is only possible to extract the form factor
$\langle  \pi(\vec 0) \pi(\vec 0) \vert J \vert 0\rangle$, with
the two pions both at rest.

The continuation of eq.~(\ref{eq:mtresult}) to Minkowski space is
obtained by replacing $t \to i t$ in the time-dependent
exponentials. At finite times therefore, the average in
eq.~(\ref{eq:pp1}) also appears in the Minkowski space version of
eq.~(\ref{eq:mtresult}). However, as shown in Appendix A, at
asymptotically large time distances $t_2$:
\begin{equation} 2E_q\,{\cal P}_{\vec q\,}(t_2)
\mathrel{\mathop{\kern0pt\longrightarrow}\limits_{t_2\to \infty }}
\frac{1}{2} \Big[\ _{\textrm{\scriptsize{out\!}}}\langle\pi(\vec
q)\pi(\meno\vec q) \,|\,J(0)\,|0\rangle\, -\,
_{\textrm{\scriptsize{in}}}\langle\pi(\vec q)\pi(\meno\vec q)
\,|\,J(0)\,|0\rangle\, \Big] \, \end{equation} so that the terms
in braces in eq.~(\ref{eq:mtresult}) combine to give the physical
amplitude (including its imaginary part). In Euclidean space this
is not the case.

\section{Finite-Volume vs Infinite-Volume Calculations:\\
Three-Dimensional Case}\label{3D}

As pointed out in sec.~\ref{sec:mt}, knowledge of $\Re\langle J
\rangle$ together with the phase shift $\delta$, is sufficient to
reconstruct the physical amplitude.  In infinite-volume the energy
spectrum is continuous and the isolation and determination of
$\rj$ turns out to be impossible. In a finite volume on the other
hand, energy levels are discrete and the extraction of $\rj$ is
possible in principle, provided that we are able to control the
preasymptotic behaviour of the relevant correlation function at
large time distances~\cite{ciuchini}.

We start this section by recalling the basic properties of the
energy levels of two-particle elastic states in a cubic periodic
box~\cite{ml}. In order to help clarify the arguments made later
in this paper, a derivation of the relevant results is given in
Appendix C.

In a finite cubic volume, $V=L^3$ (where $L$ is the length of each
spatial direction), the allowed values of the ``radial'' relative
momentum, $k$, of a two-particle cubically invariant state with
total momentum zero and with a non-zero s-wave component, satisfy
the condition
\begin{equation}
h(W,L) = n \label{phase} \, ,
\end{equation}
where $n$ is a non-negative integer~\footnote{ For $n=0$, there
are two solutions: one corresponding to $k=0$ which is spurious,
the other giving the L\"uscher relation between the finite volume
energy and the  scattering length~\cite{ml}.} and \begin{equation}
h(W,L) \equiv \frac {\phi (q) + \delta(k)}{\pi} \, . \label{hwl}
\end{equation} In eq.~(\ref{hwl}) $\delta(k)$ is the
infinite-volume s-wave phase shift, $q\equiv {{kL} / {2\pi }}$,
and $k$ is related to the center of mass energy $W$ by
\begin{equation}
    W=2\sqrt {m_\pi ^2+k^2} \, . \label{eq:w}
\end{equation}
The function $\phi(q)$ is defined by
\begin{eqnarray}
\tan \phi (q)&=&-\frac{\pi ^{\frac32}q}{Z_{00}(1;q^2)}\, ,
\\
\textrm{where}\hspace{1in}&&\nonumber\\
Z_{00}(s;q^2)&=&\frac{1}{\sqrt {4\pi}}\sum\limits_{\vec l\in Z^3}
{(\vec l^{\,\,2}-q^2)^{-s}}\label{zeta}\, .
\end{eqnarray}
In the following we will denote the energies (\ref{eq:w})
corresponding to each value of $n$ by $E_n$. Eq.~(\ref{phase}) is
derived under the simplifying assumption of an angular momentum
cutoff: here and below we therefore ignore the contribution to
eq.~(\ref{phase}) from states of higher angular momenta. These
could be included without difficulty in our discussion. Moreover
we will consider $\delta$ as a function of the total energy or of
$k$, whichever will be more convenient.

The spectrum of cubically invariant two-pion states with total
momentum zero is not exhausted by the solutions of
eq.~(\ref{phase}). There are states which correspond to $q=|\vec
l|=|\vec l^\prime|$ such that $\vec l$ and $\vec l^\prime\in Z^3$ and
are not related by a cubic rotation~\cite{ml}. It is shown in Appendix
C that these additional states do not have an s-wave component, in
contrast to those given by eq.~(\ref{phase}).

Although the energy levels are those corresponding to a finite volume
$V$, $\delta(k)$ in eq.~(\ref{hwl}) is the {\it infinite}-volume
phase shift.  The derivation of this equation~\cite{ml}, described in
Appendix C, requires that $V$ is larger than the two pion interaction
region.  In particular, in Appendix C it is shown that
eq.~(\ref{phase}) is equivalent to the condition that there is no
distortion of the s-wave component of the two-pion wave function due
to boundary conditions.  We denote these wave functions by
$\Psi_{E_{n}}^{(V)}(\vec x)$, where $\vec x$ is the relative position
of the two pions.

Our strategy is based on the analysis of correlators of local
observables which admit two pions as intermediate states.  For
simplicity we take the two pions to have total momentum zero.  We
consider local observables which are rotationally invariant in
infinite volume.  As discussed in section \ref{sec:large}, this
amounts to selecting the contribution of the s-wave component of the
state $\Psi_{E_{n}}^{(V)}(\vec x)$ to the matrix element $\langle 0 |
J(0) | E_{n} \rangle$.

In order to establish the relation between the finite and
infinite-volume amplitudes, it is convenient to consider the two-point
correlation function:
\begin{equation}
{\cal C}(t) \equiv \int\limits_V {d^{\,3}x\,\langle\,0\,|J (\vec
x,t)J(0)\,|\,0\,  \rangle }\, , \label{eq:cdef}\end{equation}
and its behaviour as the volume $V$ becomes large. We have
\begin{eqnarray}
\int\limits_V {d^{\,3}x\,\langle\, 0\,|\,J (\vec x,t)J (0)\,|\,0\,\rangle
_V}&
\mathrel{\mathop{\kern0pt\longrightarrow}\limits_{V\to \infty}} &
\frac{(2\pi )^3}{2(2\pi )^6}\int\frac{d^{\,3}p_1}{2E_1}
\frac{d^{\,3}p_2}{2E_2}\delta (\vec p_1+\vec
p_2)e^{-(E_1+E_2)t} \times\nonumber\\
&&\hspace{1.5in}
\vert\,{\langle\, 0 \,\vert\, J(0)\,\vert\,{\pi(\vec p_1\,)\,\pi(\vec p_2\,)}
\,\rangle}\, \vert^2\nonumber\\
&&\hspace{-2.3in}\rule[-0.2cm]{0cm}{1.1cm}=
\frac{1}{2(2\pi )^3}\int {dE}\,e^{-Et}\vert {\langle 0 \vert J
(0)\vert {\pi \pi ,E} \rangle } \vert^2
\int\frac{d^{\,3} p_1}
{2E_1}\frac{d^{\,3} p_2}{2E_2}\delta (\vec
p_1+\vec p_2)\delta (E-E_1-E_2)\label{uno}\\
&&\hspace{-2.3in} =\frac{\pi}{2(2\pi )^3}\int\frac{dE}{E}
e^{-Et}\vert \langle
0 \vert J (0)\vert\pi\pi, E \rangle  \vert^2 \, k(E)\ ,\nonumber
\end{eqnarray}
where $E_{1,2}=\sqrt{p_{1,2}^2+m^2}$ and
\begin{equation}
    k(E)=\sqrt {\frac{E^2}{4}-m_\pi ^2} \, .
\end{equation}
On the other hand, in a finite, but asymptotically large, volume
we have
\begin{eqnarray}
\int\limits_V {d^{\,3}x \langle\,0\,|\, J (\vec x,t) J (0)\,|\,0\rangle_V }&=&
V\sum\limits_n {\vert { \langle 0 \vert J(0)\vert {\pi \pi ,n}
\rangle _V} \vert^2e^{-E_nt}} \label{due} \\
&\mathrel{\mathop{\kern0pt\longrightarrow}\limits_{V\to \infty
}}&V\int\limits_0^\infty dE\,\rho_{V}(E)\,\vert {\langle 0 \vert J
(0)\vert {\pi \pi ,E} \rangle _V} \vert^2e^{-Et} \, .\nonumber
\end{eqnarray}

In eq.~(\ref{due}), $\vert {\pi \pi ,n} \rangle _V$ and $\vert
{\pi \pi ,E} \rangle _V$ denote the finite-volume two-pion states
at fixed $n$ and fixed energy $E$, respectively.  $\vert {\pi \pi
,n} \rangle_V$ are classified according to the quantum number $n$
defined in eq.~(\ref{phase}).  $\rho_V(E)$ denotes a function to
be determined, which provides the correspondence between finite
volume sums and infinite volume integrals. In many cases, for
example in one dimension, $\rho_V(E)$ can be identified as the
density of states at energy $E$. In section \ref{sec:large}, after
a discussion of the constraints imposed by the locality of $J(x)$,
we show that, also in the presence of cubic boundary conditions,
$\rho _{V}(E)$ is given by,
\begin{equation}
\rho _{V}(E)=\frac{dn}{dE} = \frac{q\phi^\prime(q)
+k\delta^\prime(k)}{4 \pi k^2}E\ ,\label{density}
\end{equation}
with exponential precision in the volume. The expression in
eq.~(\ref{density}) is the one  which one would heuristically
derive from a na\"{\i}ve interpretation of $\rho_V(E)$ as the
density of states, cf. eq.~(\ref{phase}). There are however, some
technical subtleties with such an interpretation which will be
discussed later in this section.

Comparing eqs.~(\ref{uno}) and (\ref{due}), we establish the
correspondence
\begin{equation}
    \vert {\pi \pi ,E} \rangle \Leftrightarrow 4\pi \, \sqrt{\frac{VE\rho_{V}
    (E)}{k(E)}}\, \vert {\pi \pi ,E} \rangle _V \, .
\label{unoonen}\end{equation}
Similarly, by considering correlators of operators which excite single
particle states, we find
\begin{equation}
\vert {\vec p=0} \rangle \Leftrightarrow \sqrt {2mV}\,\vert {\vec
p=0} \rangle _V\ .
\label{eq:duetwo}
\end{equation}
From eqs.~(\ref{unoonen}) and (\ref{eq:duetwo}) we readily obtain
the amplitudes of the effective weak Hamiltonian in terms of the
corresponding finite-volume matrix elements:
\begin{equation}
\vert {\langle {\pi \pi ,E} \vert{\cal H}_W(0)\vert K \rangle }
\vert^2=
32 \pi^{2} V^2 \, \frac{\rho_{V}(E) E m_K }{k(E)} \, \, \vert
{{}_V\langle {\pi \pi ,E} \vert{\cal H}_W(0)\vert K \rangle _V}
\vert^2 \,.\label{LLL1}
\end{equation}
In order to obtain the physical amplitude we set $E=m_K$ in the above equation
and find
\begin{equation}
\vert {\langle {\pi \pi ,E=m_K} \vert{\cal H}_W(0)\vert K \rangle
} \vert^2=
8 \pi V^2 \; \{q\phi '(q)+k\delta '(k)\}_{k=k_\pi}
\left(\frac{m_K}{k_\pi}\right)^3 \vert {{}_V\langle {\pi \pi ,E} \vert{\cal
H}_W(0)\vert K \rangle _V} \vert^2 \,
\label{LLL1b}\end{equation}
where
\begin{equation}
k_{\pi} \equiv \sqrt{\frac{m_K^2}{4}-m_\pi^2} \, .
\end{equation}
Eq.~(\ref{LLL1b}) is the same as the formula derived by Lellouch and
L\"uscher~\cite{lll}.  The additional factor of $V^2$ on the r.h.s.
of eq.~(\ref{LLL1b}) is present because we have used the Hamiltonian
density rather than the Hamiltonian.  We stress that the relation
(\ref{LLL1}) is valid for the matrix elements of any local operator
with any momentum transfer (below the inelastic threshold) and hence
the matching condition $m_K=E$ is not required.  Although
eq.~(\ref{LLL1b}) appears to be equivalent to the corresponding
equation in ref.~\cite{lll} there is an important difference in the
two derivations.  The result of ref.~\cite{lll} was obtained at a
fixed value of $n$ and at a fixed volume $V$, tuned to have
$m_K=E_{n}$, with $n < 8$.  Eq.~(\ref{LLL1}), on the other hand, has
been derived at fixed energy $E$ for asymptotically large volumes $V$.
This implies that as $V \rightarrow \infty$ we must simultaneously
allow $n \rightarrow \infty$.  A question which arises naturally at
this point is what is the relation between the two approaches?  The
answer requires a more detailed discussion, developed in the following
section, where it will be shown that the constraints of locality allow
us to establish eq.~(\ref{density}) with exponential accuracy for
elastic states under the inelastic threshold.

We end this section with a discussion of the subtleties mentioned
above, concerning a rigorous interpretation of $\rho _{V}(E)$ as a
density of states. For purposes of illustration we discuss the
issues of completeness, normalization and degeneracy for the free
theory. In this case the eigenfunctions are plane waves and we
write the completeness relation directly in the subsector of zero
total momentum
\begin{equation}
\sum_{\{\vec p_{n}\} }\frac{e^{i\vec p_{n} \cdot (\vec x -\vec
y)}}{V} = \delta (\vec x -\vec y) \hspace{2cm} \vec x, \vec y\,
\epsilon\, V \ ,\label{plane}
\end{equation}
where, as above, $\vec x$ denotes the relative position of the two
particles. Since we will consider the insertion of this
completeness relation between local, rotationally invariant
operators, the only relevant component of the plane waves is their
s-wave projection:
\begin{equation}
\psi_{p_{n}}(r) \equiv \frac{1}{\sqrt V} \int_{\Omega} \frac{d
\Omega}{4 \pi} e^{i\vec p_{n} \cdot \vec x}= \frac{1}{\sqrt V}
\frac{\sin p_{n} r}{p_{n} r}\ ,
\end{equation}
with $p_{n} \equiv \big |\vec p_{n}\big |$. The component of the
completeness relation which is equivalent to the identity when
inserted between s-wave operators reads
\begin{equation}
\sum_{\{E_{n}\}} \nu_n \psi_{p_{n}}(r) \psi_{p_{n}}(r')= {\cal
I}_{s}\,, \label{complete}
\end{equation}
where ${\cal I}_{s}$ is the s-wave projection of the identity
($\delta (\vec x -\vec y)$ appearing in eq.~(\ref{plane})) and
$\nu_n$ denotes the number of plane waves with the same $p_{n}$,
i.e.  the same energy.  The sum over states in
eq.~(\ref{complete}) is now a sum over the different energy values
$E_{n}$ rather than over the different momenta as was the case in
eq.~(\ref{plane}).

Eq.~(\ref{complete}) shows that the factor $\nu_n$ should not be
considered as a degeneracy in the s-wave projection (it
corresponds always to the same wave function); it should instead
be viewed as a normalization factor. Thus the correctly normalized
s-wave functions, when considering the completeness relation
between rotationally invariant states, are
\begin{equation}\label{ffinite}
f_{p_{n}}(r) =\sqrt \nu_{n} \psi_{p_{n}}(r) =
\sqrt{\frac{\nu_{n}}{V} } \frac{\sin p_{n} r}{p_{n} r}\ .
\end{equation}

It is now straightforward to deduce the relation between the
finite volume s-wave function, $f_{p_{n}}(r)$ and the
corresponding wave function in infinite volume:
\begin{equation}
{\psi_p}^{(\infty)}(r) \equiv \frac{1}{\sqrt {(2 \pi)^{3}}}
\int_{\Omega} \frac{d \Omega}{4 \pi} e^{i\vec p \cdot \vec x}=
\frac{1}{\sqrt{ {(2 \pi)^{3}}}} \frac{\sin p r}{p r}\ .
\label{infinite}
\end{equation}
The relation is obtained by comparing eqs.~(\ref{ffinite}) and
(\ref{infinite}):
\begin{equation}
{\psi_{p_{n}}}^{(\infty)}(r) =
\frac{1}{\sqrt{\nu_n}}\,\sqrt{\frac{V}{(2 \pi)^{3}}}\
f_{p_{n}}(r)\ .
\label{relation}\end{equation}

Recall that $\nu_{n}$ in eq.~(\ref{complete}) is the number of
different ways in which a given $\vec n ^{\,2}$ can be realized
using vectors with integer components. For large $n$, $\nu_{n}$ is
rapidly and irregularly varying with $n$ and therefore the
presence  of a factor of $\nu_n$ poses a difficulty to a rigorous
interpretation of $dn/dE$ as the density of states~\footnote{We
acknowledge many discussions on this point with L.~Lellouch and
M.~L\"uscher.}. However, what actually happens is that, whereas
both factors $\rho_{V}(E)$ and $\vert {\langle 0 \vert J (0)\vert
{\pi \pi ,E} \rangle _V} \vert^{2}$ in eq.~(\ref{due}) are rapidly
varying with energy, their product has a smooth behaviour. This
can be seen by a computation of the correlator of a rotationally
invariant observable, such as $:\phi^{2}(x):$ (where $\phi$ is a
scalar field). For $J(x)=:\phi^2(x):$ eq.~(\ref{due}) becomes
\begin{eqnarray}
& & \int_{V}d^{3} x\langle\,0\,|\,:\phi^{2}(x):\,\, :\phi^2(0):
\,|\,0\,\rangle_{V} = \label{green1}\\
& & =\frac{1}{2} \int_{V}d^{3} x \sum_{\vec
p_{1},\vec p_{2}} \big |\langle 0|
:\phi^{2}(0) :|\vec p_{1},\vec p_{2} \rangle_{V}\big
|^{2}\, \, e^{i (\vec p_{1}+\vec p_{2})\cdot \vec x}\, e^{-(
E_{p_1} + E_{p_2})t} \nonumber \\
& & = \frac{V}{2} \sum_{\vec p} \bigg |\langle 0|
:\phi^{2}(0): |\vec p,-\vec p\, \rangle_{V}\bigg |^{2}
e^{- 2 E_p t}\nonumber \\ & & = V\sum\limits_n {\vert { \langle\,
0\,\vert :\phi^{2}(0):\,\vert\, n\, \rangle _V} \vert^2e^{-E_nt}}
= \frac{1}{V} \sum_{E_{n}} \frac{2}{ E_{n}^{2}}\ \nu_{n}\ e^{- E_n
t} \nonumber \\ &
&\mathrel{\mathop{\kern0pt\longrightarrow}\limits_{V\to \infty }}
V\int\limits_0^\infty dE\ \rho_{V}(E)\ \vert {\langle\,0\,\vert
:\phi^{2}(0):\vert\, n\, \rangle _V} \vert^2\ e^{-Et}\ .\nonumber
\end{eqnarray}
In eq.~(\ref{green1}) we have used the free-field result
\begin{equation}
\langle 0| :\phi^{2}(0): |\vec p,-\vec p \rangle_{V}=\frac{2}{V
E_{n}}\,,
\end{equation}
so that
\begin{equation}
| \langle\, 0\,| :\phi^{2}(0) : |\,n\,\rangle |^{2}=
\frac{2}{V^{2} E_{n}^{2}}\,\nu_{n}\ .
\end{equation}
Therefore the correspondence in eq.~(\ref{unoonen}) is given in
the free theory by
\begin{equation}\label{eq:ffrel}
\big |\,\langle\,0\,| :\phi^{2}(0) :|\,\vec p_{1},\vec
p_{2}\,\rangle\,\big |^{2} = |\, \langle\, 0\, | :\phi^{2}(0) :
|\, n\, \rangle\, |^{2} \, \,  E_{n}^{2} \,
\frac{\phi^\prime(n)}{2 \pi^{2} n^{2}} V^{2}\ .
\end{equation}
In deriving eq.~(\ref{eq:ffrel}) we have used the fact that in the
free case the spectrum of states, and hence the solution of
eq.~(\ref{phase}), is given by
\begin{equation}
q^{2} =\bigg (\frac{L k}{2 \pi} \bigg )^{2} = n^{2}\ .
\label{freequant}
\end{equation}
It is easy to demonstrate that
\begin{equation}
\phi^\prime(n) = \frac{4 \pi^{2} n^{2}}{\nu_{n}}\ ,
\end{equation}
and hence that
\begin{equation}
\big |\langle\, 0\,| :\phi^{2}(0) :|\,\vec p_{1},\vec p_{2}\,
\rangle\big |^{2}= 4\ ,
\end{equation}
which is the correct result for the free-field theory with
covariant normalization in infinite volume.

Following this discussion, the last line of eq.~(\ref{green1}) is
equal to
\begin{equation}
V\int\limits_0^\infty dE\,\rho_{V}(E)\,\vert {\langle\,0\,\vert
:\phi^{2}(0):\vert\, n\, \rangle _V} \vert^2\ e^{-Et} = \frac{1}{4
\pi^2}\int_{2 m}^{\infty} d E\ \frac {k(E)}{E}\ e^{-E t}
\end{equation}
in perfect agreement with the infinite volume expression.

This example shows that, in the free case, the meaning of the LL
factor is very simple.  It removes the degeneracy factor ($\nu_{n}$)
due to the fact that we extract from the correlator the contribution
of the $n^{\textrm{th}}$ energy level and restores the appropriate
factors to normalize the states in the infinite volume limit
correctly.

\section{How large is a \textit{large volume}?}
\label{sec:large}

In this section we study finite-volume effects in a toy
mathematical example as well as in non-relativistic quantum
mechanics and relativistic field theory.  The common feature and
key ingredient in all these cases is the locality of the operators
whose matrix elements are being computed.  Within our approach we
will be able to show that, in the quantum mechanical case, the LL
formula is true for all levels containing an s-wave component, up
to exponentially small corrections in the volume.  This conclusion
is compatible with a stronger unpublished result by
L\"uscher~\cite{notes}, who finds that the LL relation is indeed
{\it exactly} valid for the same levels.  The discussion given in
ref.~\cite{notes}, however, is specific to non-relativistic
quantum mechanics and cannot be readily generalized to
relativistic quantum field theory.

In relativistic field theory the situation is more complicated for two
basic reasons.  Firstly, the interaction range is not precisely
defined, so that all results are valid at most with exponential
precision.  The second and more substantial difficulty comes from the
presence of inelastic production, which, as shown in
sec.~\ref{sec:wave} below, could also affect the validity of the
quantization condition in eq.~(\ref{phase}).

For the discussion below, it will be useful to recall the general
relation between a discrete sum and the corresponding integral, which
follows from the Poisson formula~\cite{light}
\begin{equation}
\sum_{n=-\infty}^{\infty} \delta(x- n) =\sum_{n=-\infty}^{\infty}
e^{i 2\pi n x}\ . \label{poisson}
\end{equation}
Multiplying both sides of eq.~(\ref{poisson}) by a function $f(x)$
and integrating over $x$, we get:
\begin{eqnarray}
\sum_{n=-\infty}^{\infty} f(n) &=& \sum_{n=-\infty}^{\infty}
\int_{-\infty}^{\infty} f(x)\, e^{2\pi i n x} dx  \label{sum}\\ &
= & \int_{-\infty}^{\infty} f(x)\, dx +\sum_{n\neq 0}
\int_{-\infty}^{\infty} f(x)\, e^{2\pi i n x}\, dx\ .\nonumber
\end{eqnarray}
Eq.~(\ref{sum}) shows that an integral is equal to a discrete sum
up to correction terms, given by the sum over terms with $n\neq
0$. In the following we will show how to control these corrections
in order to make contact with the LL formula.

\subsection{A Toy Example}
Consider the simple one-dimensional example of a
$C^{\infty}$-function, $f(x)$, vanishing with all its derivatives
outside a compact support which is entirely contained inside the
``finite volume'' $-L/2 <x < L/2$.  This is the statement of
locality in our toy model.  $f(x)$ can be expanded as a Fourier
series:
\begin{equation}
    f(x) = \frac{1}{L} \sum_{n} \tilde {f}(p_{n}) e^{ip_{n}x} \, ,
    \label{fourier}
\end{equation}
where
\begin{equation}
p_{n}\equiv\frac{2\pi}{L}\,n
\label{discrete}\end{equation}
and
\begin{equation}  \tilde {f}(p) \equiv \int_{-L/2}^{{L/2}}dx \,
    f(x) e^{-ipx} =  \int_{-\infty}^{{+\infty}}dx \,  f(x) e^{-ipx} \, .
\end{equation}
On the other hand, $f(x)$ can also be written as a Fourier
transform
\begin{equation}
f(x) = \frac{1}{2\pi} \int_{-\infty}^{+\infty}dp \, \tilde {f}(p) e^{ipx} \, .
\label{transform}\end{equation}
Whereas eq.~(\ref{transform}) reproduces $f(x)$ exactly,
eq.~(\ref{fourier}) coincides with $f(x)$ only in the interval $-L/2 <x
< L/2$; outside this interval $f(x)$ is replicated periodically on the
$x$-axis. If we now consider a second function $g(x)$ with the same
properties as $f(x)$, we have:
\begin{equation}
\int_{-\infty}^{+\infty}dx \, g^*(x)f(x)=\frac{1}{L}\sum_{n}
\tilde {g}^{*}(p_{n}) \tilde {f}(p_{n})=\frac{1}{2
\pi}\int_{-\infty}^{+\infty} dp \, \tilde {g}^{*}(p) \tilde
{f}(p)\,. \label{dictionary}
\end{equation}
Eq.~(\ref{dictionary}) is an example of the exact equality between
an integral and a discrete sum without any correction terms.  It
shows that, for local observables, the ``finite volume
correlator'' $\frac{1}{L}\sum_{n} \tilde {g}^{*}(p_{n}) \tilde
{f}(p_{n})$ is exactly equal to the ``infinite volume correlator''
$\frac{1}{2 \pi}\int_{-\infty}^{+\infty} dp \, \tilde {g}^{*}(p)
\tilde {f}(p)$, as long as the support of the observables is
entirely contained inside the ``quantization volume'' $L$.  In the
next section we will show how to derive similar relations to the
LL-formula from equations such as (\ref{dictionary}).

\subsection{Finite-Volume Effects in Non-Relativistic Quantum
Mechanics} \label{QM}

We now extend the previous considerations to the case of a quantum
system, starting with the non-relativistic quantum mechanical
case. In preparation for the relativistic case presented below, we
use the formalism of second quantization also in the quantum
mechanical case and introduce interpolating operators
$\phi^\dagger(\vec x,t)$ and $\phi(\vec x,t)$ that respectively
create and annihilate the physical particles. Consider two
``local'', zero angular-momentum two-particle operators ${\cal
R}(\vec X,t)$ and ${\cal S}(\vec X,t)$ defined by
\begin{equation}
{\cal R}(\vec X,t)=\frac12\int d^{\,3}
x\,R(r)\,\phi^{\dagger}\left(\vec X+\frac{\vec
x}{2},t\right)\phi^{\dagger}\left(\vec X-\frac{\vec x}{2} , t
\right) \, ,\quad \mbox{where} \quad r\equiv|\vec x| \, ,
\label{local}
\end{equation}
with a similar definition for ${\cal S}(\vec X, t)$. The functions
$R(r)$ and $S(r)$ defined in eq.~(\ref{local}) have support
contained inside $V$, thus ensuring locality. Under this
hypothesis, we now show that the matrix elements of ${\cal R}$ and
${\cal S}$ in the finite volume $V$ are proportional to those in
the infinite volume, with a constant of proportionality given
precisely by the LL-factor.

Consider the zero momentum two-point function
\begin{eqnarray}
{\cal G}_V(t)\equiv\int_V d^{\,3} X\,\langle\,0\,|\,{\cal
S}^\dagger(\vec X,t) \,{\cal R}(\vec 0,0 )\, |\,0\,\rangle \ .
\label{equal}
\end{eqnarray}
At $t=0$ we have
\begin{equation}
{\cal G}_V(0) = \int_{V} d^{3} x S(r) R(r) = \int d^{3} x S(r)
R(r) = {\cal G}_{\infty}(0)\,, \label{inffin}
\end{equation}
where the second equality comes from the hypothesis that $S(r)$
and $ R(r)$ have support only in $V$. Eq.~(\ref{inffin}) shows
that we can analyze ${\cal G}_V(0)$ both in terms of finite volume
and infinite volume intermediate states.  Inserting a set of
complete states in the infinite volume we have
\begin{eqnarray}
{\cal G}_{\infty}(0) = \int _{\bar E}^{\infty} dE\,s^{*}(E)\,
r(E) \, \label{continuous}
\end{eqnarray}
where $\bar E$ is the lowest energy in the $\pi \pi$ channel,
$\bar E=0$ or $2 m_{\pi}$ in the nonrelativistic and relativistic
cases respectively and, for example,
\begin{equation}
s^{*}(E)= \int d^{3} x S(r) \Psi_{E}^{(s-wave)}(r) \, .
\label{matrix}
\end{equation}
In eq.~(\ref{matrix}) we have denoted by $\Psi_{E}^{(s-wave)}(r)$
the center of mass, zero-angular momentum wave function of two
pions with energy $E$.

We now turn to the correlation function in a finite volume, and
insert a complete set of energy eigenstates:
\begin{equation}
{\cal G}_V(0) = V \, \sum_{n} \langle\,0\,|{\cal S}^\dagger(\vec
0, 0)|n\rangle\langle n| {\cal R}(\vec 0,0)|\,0\,\rangle\ ,
\label{finite}
\end{equation}
where $|n\rangle$ denote the energy eigenstates with zero total
momentum and with the normalization $\langle n\vert
n^\prime\rangle=\delta_{n,n^\prime}$. Since the local operators
under consideration depend  only on $r$, the sum in
eq.~(\ref{finite}) receives contributions only from the zero
angular momentum component of the intermediate states. In other
words we have, for example
\begin{eqnarray}
\langle 0 | {\cal S}^\dagger(\vec 0, 0) | n \rangle = \int_{V} d^{\,3}
x \, S(r) \Psi_{E_{n}}^{(V)}(\vec x) \propto s^{*}(E_{n}) \, .
\end{eqnarray}
As shown in Appendix C, the states $|n\rangle$ are precisely those
described by eq.~(\ref{phase}).  We may therefore rewrite
eq.~(\ref{finite}) as
\begin{equation}
{\cal G}_V(0) = V \sum_{n} s^{*}(E_{n}) r(E_{n})
c_{V}(E_{n}) \,  ,
\label{time}
\end{equation}
with a set of non-negative coefficients $c_{V}(E_{n})$, in terms
of which we have
\begin{equation}
| s(E_{n}) |^{2}=\frac {1}{c_{V}(E_{n})} | \langle 0 | {\cal
S}^\dagger(\vec 0, 0) | n \rangle |^{2}\ . \label{norma}
\end{equation}
It is our aim to evaluate the coefficients $c_{V}(E_{n})$ and to
demonstrate that eq.~(\ref{norma}) is precisely the LL-relation.

The previous discussion shows that at $t=0$ the correlation
function ${\cal G}_V(0)$ defined in eq.~(\ref{equal}) coincides
with the corresponding one in the infinite-volume limit, without
any corrections. We now investigate what happens at a generic
value of $t$. In this case we get
\begin{equation}
{\cal G}_V(t) = V \, \sum_{n} \langle\,0\,|{\cal S}^\dagger(\vec 0,
0)|n\rangle\langle n| {\cal R}(\vec 0,0)|\,0\,\rangle e^{-E_{n} t} =
\int_{V} d^{3} x S(r,t) R(r)
\label{finitet}
\end{equation}
where
\begin{eqnarray}
S(r, t) \equiv  e^{-Ht} S(r)\, ,
\label{eq:mostra}
\end{eqnarray}
with a similar result for $R(r,t)$.  In eq.~(\ref{eq:mostra}) $H$
denotes the first quantization, center of mass Hamiltonian
(differential) operator acting on two-particle wave functions,
whose eigenvalues are precisely the $\{E_{n}\}$. From
eq.~(\ref{eq:mostra}) it follows that $S(r, 0)= S(r)$ and $R(r,
0)=R(r)$.

The key point for the identity of finite and infinite volume
correlators at $t =0$, as seen from eq.~(\ref{inffin}), is the
locality
of $S(r)$ and $R(r)$.  On the other hand, for a generic $t$,
eq.~(\ref{eq:mostra}) shows that $S(r, t)$ is obtained from $S(r)$
through a diffusion process.  It is well known that, even if
$S(r)$ has a compact support, $S(r,t)$ will be non-zero over the
entire infinite volume for $t\neq 0$.  In general, therefore
\begin{equation}
{\cal G}_V(t) \neq {\cal G}_\infty(t)\ .
\end{equation}
However, diffusion is a ``slow'' process, and for a range of $t
\approx L$ we still have
\begin{equation}
{\cal G}_V(t)={\cal G}_{\infty}(t) + {\cal O} (e^{-m_{\pi}L})\ ,
\label{green}
\end{equation}
while for $t > m_{\pi} L^{2}$ we expect ${\cal G}_V(t)$ to be very
different from ${\cal G}_{\infty}(t)$.  Eq.~(\ref{green}) implies
that, with exponential accuracy in the volume,
\begin{equation}
V \sum_{n} s^{*}(E_{n}) r(E_{n}) c_{V}(E_{n}) e^{-E_{n} t} =
\int_{\bar E}^{\infty} d E s^{*}(E) r(E) e^{-E t}\ .\label{contin}
\end{equation}
We can now use eq.~(\ref{poisson}) to transform the r.h.s. of
eq.~(\ref{contin}) back into a sum again. Eq.~(\ref{poisson})
implies that
\begin{equation}
\sum_{n=-\infty}^{\infty} \delta(h(E,L)-n) =
\sum_{n=-\infty}^{\infty} e^{i 2 \pi n h(E,L)}\ , \label{sumq}
\end{equation}
where $h(E,L)$ is defined in eq.~(\ref{hwl}).

Combining eqs.~(\ref{contin}) and (\ref{sumq}) we obtain
\begin{equation}
\int _{\bar E}^{\infty} d E s^{*}(E) r(E) e^{-E t} = \sum_n \frac
{s^{*}(E_{n}) r(E_{n}) e^{-E_{n} t}}{\big |\frac {\partial
h(E,L)}{\partial E}\big |_{E_{n}}} - {\cal Q}(L,t)\ ,\label{remaind}
\end{equation}
where the observable ${\cal S}$ and ${\cal R}$ have been chosen to
vanish sufficiently rapidly in $\bar E$ to avoid problems with the
spurious solution with $k=0$ (see below). In eq.~(\ref{remaind}),
we have introduced
\begin{equation}
{\cal Q}(L,t) \equiv \sum_{l \neq 0} \int _{\bar E}^{\infty} d E
s^{*}(E) r(E) e^{-E t} e^{i 2 \pi l h(E,L)} \label{eq:q}
\end{equation}
and used the fact that the energies are non-negative.

A general property of any quantization condition is that for fixed
$E$
\begin{equation}
h(E,L) \mathrel{\mathop{\kern0pt\longrightarrow}\limits_{L\to
\infty }} \infty\ .\label{infinite1}
\end{equation}
In our case, with cubic boundary conditions, we have
\begin{equation}
h(E,L) \mathrel{\mathop{\kern0pt\longrightarrow}\limits_{L\to
\infty }} L \, k(E) \, . \end{equation}
Eq.~(\ref{infinite1}) is a statement that as the volume is
increased, the value of $n$ corresponding to any fixed energy also
increases.

For $n=0$, a possible complication may arise due to the spurious
solution with $k=0$. This can be easily avoided however, since it
is possible to choose local operators for which the threshold
behaviour of $s^{*}(E)\,r(E)$ can be made to vanish arbitrarily
quickly. This can be achieved by taking appropriate combinations
of local observables and their time derivatives and recalling that
any time derivative multiplies the matrix element by the energy
according to
\begin{equation}
\langle 0 | {\cal {\dot S}}^\dagger(\vec 0, 0) | n \rangle = i
\langle 0 | [H,{\cal S}^\dagger(\vec 0, 0)] | n \rangle = - i E
\langle 0 | {\cal S}^\dagger(\vec 0, 0) | n \rangle \ .
\end{equation}

Using the techniques of asymptotic analysis~\cite{bleis}, and taking
into account eq.~(\ref{infinite1}), we deduce that the behaviour of
the correction term ${\cal Q}(L,t)$ at large $L$ is dominated by the
critical points of $h(E,L)$, i.e.  the points at which $h(E,L)$ has a
vanishing derivative or the points where $h(E,L)$ and $s^{*}(E)\,r(E)$
are not continuous or differentiable.  In the present case the only
critical point of $h(E,L)$ occurs at threshold, i.e.  for $k \approx
0$.  This means that the behaviour of ${\cal Q}(L,t)$ at large $L$ is
power-like, with the power determined by the threshold behaviour of
$s^{*}(E) r(E)\, e^{-E t}$. These power corrections depend on the time
as $e^{-\bar E t} \sim e^{-2 m_\pi t}$ (multiplied by powers of $t$).
They therefore only affect matrix elements computed at energies close
to threshold. Indeed, as discussed above, by an appropriate choice of
local observables it is possible to make the product $s^{*}(E) r(E)$
at threshold arbitrarily small and consequently to make the leading
power in the $1/L$ expansion in ${\cal Q}$ arbitrarily high. Thus 
\begin{equation}
V \sum_{n} s^{*}(E_{n}) r(E_{n}) c_{V}(E_{n}) e^{-E_{n} t}= \sum_n
\frac {s^{*}(E_{n}) r(E_{n}) e^{-E_{n} t}}{\big |\frac {\partial
h(E,L)}{\partial E}\big |_{E_{n}}}  - {\cal Q}(L,t)\, ,
\end{equation}
and since we are
interested in amplitudes at fixed energy $E$ (between $\bar E$ and the
inelastic threshold), in the limit $L \to \infty$, $n \to \infty$,
we have, with exponential accuracy,
\begin{equation}
c_{V}(E_{n}) = \frac{1} {V  \big |\frac {\partial
h(E,L)}{\partial E}\big |_{E_{n}}} \label{accuracy}
\end{equation}
for all states with energies above $\bar E$.
Eqs.~(\ref{norma}) and~(\ref{accuracy}), together with the kinematic
factors appropriate to the chosen normalization, establish the LL-type
relations of the kind given in eq.~(\ref{LLL1}), with exponential
accuracy.

We conclude this subsection with some brief remarks about the
matrix element with the two-pion state at threshold. A priori, the
threshold matrix element can be affected by power corrections,
although our derivation does not necessarily imply that this is
the case, since we are unable to control the sum over $l$ in
eq.~(\ref{eq:q}). Threshold states are generally problematic and
require a separate discussion. For example, in the case of an
attractive potential, their energy in a finite volume is lower
than $\bar E$. We are not able therefore, to derive a relation
between the finite-volume matrix element at threshold and the
corresponding  infinite-volume amplitude which we can demonstrate
is free of corrections which vanish as inverse powers of the
volume.

\subsection{Finite-Volume Effects in Relativistic Quantum Field Theory}
In relativistic quantum field theory (RQFT) the previous
considerations must be modified since strict localization of the
support is not possible and therefore the state
\begin{equation}
    | J\,\rangle\equiv J(0)|\,0\,\rangle
\end{equation}
is only approximately localized around the origin.  The
probability of finding particles at a distance $r$ away from the
origin never vanishes, although, as a consequence of clustering,
it decreases exponentially like $e^{-m_{\pi} r}$~\cite{thirring}.
There is a further point that needs to be considered.  The
correlator appearing in eq.~(\ref{inffin}) is evaluated at equal
times. In field theory such an equal-time correlator is generally
UV divergent, due to the short distance singularities in the
product of two local operators. A natural regulator for such
divergences is the (Euclidean) time. The discussion of the quantum
mechanical case can then be generalized to field theory. We
therefore conclude that the equality
\begin{equation}
{\cal G}_V(t)={\cal G}_{\infty}(t)
\end{equation}
is also valid in RQFT with exponential accuracy.

For correlators of local operators, the most important difference
from the non-relativistic quantum mechanical case is due to the
existence of the inelastic threshold, $E_{th}$. This implies that
in any fixed finite volume, there are only a limited number of
elastic states with energies below $E_{th}$. Therefore the
truncated sum over the $E_{n}$ (truncated at $E_{th}$) reproduces
the corresponding integral (cut-off at the same energy) only up to
corrections which vanish as $L \to \infty$. However, the number of
such states increases with the volume and the arguments presented
in section \ref{QM} can readily be generalized to the RQFT. In
particular we conclude that eq.~(\ref{accuracy}) is still valid
with exponential accuracy. Our derivation of the LL formula
demonstrates that, in principle, it is possible to improve the
precision by working at fixed energy on increasing volumes (and
hence increasing $n$).

The discussion above does not imply, however, that the corrections are
small for any volume.  In order for these corrections to be really
negligible, we must work in volumes which are sufficiently large that
the sum over energies is well approximated by the corresponding
integral.  It is of course impossible to give an estimate of the error
on the LL formula in lattice simulations when the volume is such that
there are only a few elastic states ($2\div 3$) below $E_{th}$.  We
will argue in the next section that in such cases there are related
difficulties in the derivation of the quantization condition
(\ref{phase}).

\section{Quantization of Two-Particle Energy Levels in
\\ Quantum Field Theory} \label{sec:wave}

In this section, we re-examine the quantization condition given in
eq.~(\ref{phase}) within the framework of Relativistic Quantum Field
Theory (RQFT).  Although eq.~(\ref{phase}) has been discussed in
detail in the quantum mechanical case, and its extension to RQFT can
also be found in ref.~\cite{ml}, to our knowledge the explicit,
non-perturbative treatment given below is new.  This discussion helps
to clarify what happens when only few elastic levels exist below
$E_{th}$.

The concept of a wave function in RQFT is an approximate one. In
this case, the object closest to a wave function is the
Bethe-Salpeter (BS) wave function. We discuss the infinite-volume
case  first. For an incoming state with total momentum zero, the
BS wave function is defined as
\begin{equation}
\Phi_{\vec p\,}(x)=\langle\, 0\,|\,T[\,\phi(\vec x, t)\, \phi(0)\,]\,
\state\ .
\end{equation}
For simplicity we present our discussion for $t=0$~\footnote{There are
no ultraviolet problems, even if $\phi(x)$ is a composite field (as is
the case in QCD) because we always choose $\vec x \neq \vec 0$.}. We
now derive the relation between $\Phi_{\vec p\,}(\vec x)\equiv\Phi_{\vec
p\,}(\vec x, 0)$ and the scattering phase shifts in the infinite-volume
limit. We introduce a complete set of intermediate states~\footnote{In
order to simplify the presentation we will treat the particles as
distinguishable.}
\begin{eqnarray}
\Phi_{\vec p\,}(\vec x) & = &\sum_{n}\langle 0 |\phi(\vec x,0) |n
\rangle \langle n| \phi(0) \state\\ &=&\frac{\sqrt{Z}}{(2
\pi)^3}\int \frac{d^{\,3}q}{2 E_{q}}\ \langle \pi(\vec q\,)\,|
\phi(0) \state \, e^{i\vec q \cdot \vec x} + {\cal I}(\vec x)\ ,
\label{eq:bs}
\end{eqnarray}
where ${\cal I}(\vec x)$ represents the contribution coming from
inelastic intermediate states with more than one particle. We
separate the connected and disconnected contributions to
eq.~(\ref{eq:bs}) in the standard way:
\begin{equation}
\Phi_{\vec p\,}(\vec x)= Z\,e^{i\vec p \cdot \vec x}+
\frac{\sqrt{Z}}{(2\pi)^3}\,\int\frac{d^{\,3} q}{2 E_{q}}\,
\langle\,\pi(\vec q\,)\,|\,\phi(0)\,\state^{\textrm{conn}} e^{i\vec q
\cdot \vec x} + {\cal I}(\vec x) \,.
\label{eq:wf}\end{equation}
Using
\begin{equation}
\langle\vec q\,|\phi(0)\,\state^{\textrm{conn}} =-\frac{\sqrt{Z} \,
{\cal M} }{q^2-m_\pi^2 +i\varepsilon}= \frac{\sqrt{Z}} { 4
E_p}\frac{{\cal M}}{ E_q-E_p - i\varepsilon}\ ,
\label{amplitude}
\end{equation}
we obtain
\begin{eqnarray}
\Phi_{\vec p\,}(\vec x)= Z \, \left(  e^{i\vec p \cdot \vec x}
+\,\frac{1}{(2\pi)^3}\int\frac{d^{\,3} q}{2 E_{q}}\, \frac{{\cal
M}}{4 E_p (E_q-E_p - i\varepsilon)} e^{i\vec q \cdot \vec x}
\right)  + {\cal I}(\vec x) \, .
\label{eq:wf1}
\end{eqnarray}
For large values of $\vec x$ the inelastic contribution can be
safely neglected. Moreover, in eq.~(\ref{eq:wf1}), the integral
over $\vec q$ is dominated by the on-shell contribution ($E_q
\approx E_p$), for which ${\cal M}$ becomes the physical on-shell
amplitude ${\cal M}(\vec p \to \vec q\,)$. Indeed ${\cal M}$ has
no singularities for $2E_p$ below the inelastic threshold. To
simplify the discussion, we consider a situation in which all the
phase shifts except the s-wave one (denoted by $\delta(p)$) are
negligible. Under this hypothesis we have
\begin{eqnarray}
{\cal M}(\vec p \to \vec q\,) = \frac{4 \pi}{i} \frac{2 E_p}{p}
\left( e^{2 i \delta(p)}-1\right) \, .\label{eq:ondas}
\end{eqnarray}
From eq.~(\ref{eq:wf1}), we then get
\begin{eqnarray}
\Phi_{\vec p\,}(\vec x)= Z \, \left( e^{i\vec p \cdot \vec x}+
\frac{1}{2 i}  \left(e^{2 i \delta(p)}-1\right) \frac{ e^{i p r}}  {p r}
\right) + \dots \, .
\label{eq:bellu}\end{eqnarray}
Using the asymptotic behaviour of the s-wave projection of $e^{i\vec p
\cdot \vec x}$
\begin{eqnarray}
e^{i\vec p \cdot \vec x} \left |_{s-wave} \right .  \sim \frac{ \sin
 p r} {p r} + \dots \, , \label{eq:bel}
\end{eqnarray}
we finally arrive at
\begin{eqnarray}
\Phi_{\vec p}(\vec x)\left |_{s-\textrm{wave}} \right . &=&
\frac{Z}{ p r}
 \left( \sin  p r +
\frac{1}{2 i} \left(e^{2 i \delta( p)}-1\right) e^{i  p r}
\right) \nonumber \\ &=& \frac{Z e^{ i \delta( p)}}{  p
r} \sin \left( \delta( p)+  p r\right) +\dots \, . \label{eq:asy}
\end{eqnarray}

We are now in a position to discuss the quantization of energy
levels in a finite volume. For our purposes it is sufficient to
consider the simple case of a spherical finite volume of radius
$R$ with vanishing fields on the boundary~\footnote{We return to
consider the quantization condition in a cubic volume in some
detail in Appendix C below.}.
For sufficiently large $R$, the asymptotic expression in
eq.~(\ref{eq:asy}) to hold, we obtain the quantization condition
\begin{equation}
n \pi - \delta(p)  = p R\ ,
\label{phase1}\end{equation}
which is the analog of eq.~(\ref{phase}) (which was derived for a cube)
on a spherical volume.

The central question for our discussion is whether the
quantization condition in eq.~(\ref{phase1}) holds when the volume
is so small that only a few levels exist below $E_{th}$. In such a
situation we show that in general it is not possible to follow the
steps that lead from eq.~(\ref{eq:wf}) to (\ref{phase1}) from
which the quantization condition is obtained.

As previously discussed, the validity of the L\"uscher
quantization condition requires (the s-wave projection of)
$\Phi_{\vec p\,}(\vec x)$ to be undeformed by the presence of the
boundary.  This implies that, up to terms which vanish
exponentially with the volume, it is
the same function of $\vec x$ (for $\vec x \epsilon V$) when
expressed as a sum over either the finite or the infinite-volume
energy eigenstates
\begin{eqnarray}
\Phi_{\vec p}(\vec x) &=& Z \, \left(  e^{i\vec p \cdot \vec x}
+  \,\int\frac{d^{\,3} q}{(2\pi)^3 2 E_{q}}\,
\frac{{\cal M}}{4 E_p (E_q-E_p - i\varepsilon)}
 e^{i\vec q \cdot \vec x} \right)  + {\cal I}(\vec x)\nonumber \\
& = &\sum_{n}\langle 0 |\phi(0) |n \rangle \langle n| \phi(0) \state
\,e^{i\vec q_{n} \cdot \vec x} \, ,
\label{eq:nostro1}\end{eqnarray}
where $\vec q_{n}$ are the $3$-momenta appropriate to the given
boundary conditions.

Both the $\delta$-function and the principal part of the pole at
$E_q=E_p$ in eq.~(\ref{eq:wf1}) are essential to obtain
eq.~(\ref{eq:bellu}); they separately contribute to both the
outgoing and incoming waves and it is only in their sum that the
principal part cancels the incoming wave and doubles the outgoing
one.  It follows that the scattered wave is purely outgoing, as
expected.  We remind the reader that for this derivation of the
quantization condition to be valid it is necessary to work at
distances such that the integral is dominated by energies $E_q
\simeq E_p$, so that the variation of ${\cal M}$ with the energy
can be neglected.  Thus a necessary condition for the validity of
the quantization formula in a finite volume is that the same
situation is reproduced by the discrete sum in
eq.~(\ref{eq:nostro1}).
Strictly speaking, in a fixed volume, it is not possible to
separate the connected and disconnected terms and to parametrize
the amplitude as in eq.~(\ref{amplitude}), because no energy poles
can be present in this case.
These contributions and, in particular, the principal part of the
integral in (\ref{eq:wf1}) must be well approximated by the sum
over those elastic states for which we may neglect the variation
of ${\cal M}$. Therefore, for the validity of eq.~(\ref{phase1}),
the sum over $n$ in eq.~(\ref{eq:nostro1}) must be dominated by
its elastic part.  On the other hand, we expect that, even
restricting the sum to one-particle intermediate states ($| n
\rangle = |\vec q_{n}\rangle$), the matrix element $\langle \vec
q_{n} | \phi(0) \state$ receives important contributions from the
inelastic states when $E_{n}>E_{th}$.  This happens because there
is no reason for the off-shell amplitude $\langle n|\phi(0)\state$
to be dominated by the one particle term. We conclude, therefore,
that in the case where only a few levels have energies below
$E_{th}$, the sum over the finite-volume eigenstates cannot, in
general, be dominated by the elastic contribution, and we should
expect corrections to eq.~(\ref{phase1}).

An equivalent way of arriving at the same conclusion is to write
the Schr\"odinger-like eigenvalue equation which is satisfied by
$\Phi_{\vec p\,}(\vec x)$, namely
\begin{eqnarray}
\left( m_\pi^2 - \nabla^2 - E^2_p \right)\Phi_{\vec p\,}(\vec x) =
\int\frac{d^{\,3} q}{(2\pi)^3 2 E_{q}}\, {\cal M} e^{i\vec q \cdot \vec x}
+ {\cal K}(\vec x) \equiv {\cal K}_{el}(\vec x)+ {\cal K}(\vec x)\, ,
\label{eq:sc}
\end{eqnarray}
where ${\cal K}(\vec x)=\left( m_\pi^2 - \nabla^2 - E^2_p
\right){\cal I}(\vec x)$.  In the non-relativistic case the r.h.s.
of eq.~(\ref{eq:sc}) is simply $V(\vec x) \Phi_{\vec p}(\vec x)$
and is therefore well localized, as discussed in the previous
section.  In the field theoretical case, ${\cal K}_{el}(\vec
x)+{\cal K}(\vec x)$ is an exponentially localized function in
$\vec x$, since it is the Fourier transform of a regular function.
The differential operator $\left( m_\pi^2 - \nabla^2 - E^2_p
\right)$ applied to $\Phi_{\vec p}(\vec x)$ filters out the long
distance contribution from the disconnected and pole terms,
leaving only the short range contribution. Thus, for the
quantization condition to be valid ${\cal K}_{el}(\vec x)+ {\cal
K}(\vec x)$ must be completely localized inside the quantization
volume.  Once again this is true if and only if ${\cal
K}_{el}(\vec x)+ {\cal K}(\vec x)$ is the same function, when
expressed as a Fourier integral or a Fourier series. However,
while in the infinite-volume case there is a substantial
contribution to ${\cal K}_{el}(\vec x)+ {\cal K}(\vec x)$ coming
from elastic states with energies below $E_{th}$, in a finite
volume this contribution is limited to a sum over a few (perhaps
even as few as two or three) elastic states and the Fourier series
and integral cannot be expected to be identical. It is true that,
in the sum over intermediate states, we still have an infinite sum
due to the contribution of all the states above $E_{th}$, however,
in cases in which inelasticity is important, this contribution is
qualitatively different from the ones below $E_{th}$.

We end this section by noting that there is a favourable dynamical
situation in which the quantization conditions (\ref{phase}) (and
(\ref{phase1})) remain valid and the LL correction formula is
applicable  even on volumes with only a few states below $E_{th}$.
This is the case for weakly interacting particles, for which
inelastic contributions are negligible at the energy of interest.
Such a dynamical situation is effectively the same as
in quantum mechanics, where it is sufficient for the interaction
region to be only slightly smaller than the quantization volume.
From phenomenological studies it is likely that such a situation
describes, to a good approximation, the low energy dynamics of
pions~\cite{daf}.

\section{Evaluation of the amplitudes: a different method}
\label{sec:delta}

In this section we analyze a different way to obtain weak $K\to\pi\pi$
amplitudes in finite-volume simulations, one that is more closely
related to the discussion in ref.~\cite{mt} and section~\ref{sec:mt}.
The method, which is currently used in numerical simulations, 
requires the evaluation of four-point
correlation functions of the form $\langle\,0\,|\pi_{\vec
q}(t_1)\pi_{-\vec q}(t_2){\cal H}_{W}(0) K(t_K)\,|\,0\, \rangle$,
where the times $t_{1,2}$ are large and positive with $t_1\gg t_2$ and
$t_K$ is large and negative. We will show that from this correlation 
function one obtains directly the real part of the decay amplitude,
up to corrections which vanish as inverse powers of the volume and
which cannot be removed by a multiplicative correction
factor. The connection of this calculation with the LL approach   
will be explained.

Since single-particle states do not present any theoretical
complications, we proceed as in section~\ref{sec:mt}, eliminating the
kaon and considering the correlation function ${\cal G}$ defined in
eq.~(\ref{eq:gdef}).When $t_1\to\infty$, eq.~(\ref{eq:gdef}) can be
conveniently written as
\begin{equation}
{\cal G}(t_1,t_2;\vec q\,) \to \frac{\sqrt{Z}}{2E_q}\,e^{-E_qt_1}\,
{\cal G}_3(t_2,\vec q\,) \, ,
\end{equation}
where
\begin{equation}
{\cal G}_3(t,\vec q)\equiv\langle\pi(\vec q\,)\,| \Phi_{-\vec
q\,}(t)\,J(0)\,|\,0\rangle \, ,
\label{eq:g3defb}\end{equation}
for $t>0$.
 From eqs.~(\ref{eq:mtresult})--(\ref{eq:pp1}) we obtain
\begin{eqnarray}
& & {\cal G}_3(t, \vec q\,)= \frac{\sqrt{Z}}{2E_q}\, e^{-E_qt}\
\big|\,_{\textrm{out}}\langle\,\pi(\vec q\,) \pi(-\vec
q\,)\,|\,J(0)\,|\,0\,\rangle\,\big|\,\cos \delta(2 E_q)\,+\cdots\ ,
\label{eq:cosdelta}\end{eqnarray}
where the ellipses represent the principal value term defined in
eq.~(\ref{eq:pp}). We remind the reader that the above equations 
are valid in infinite volume.

\par If we could isolate the term proportional to $\exp(- E_q\,t)$ in
the correlation function ${\cal G}_3$, we would obtain
$\vert_{\textrm{out}}\langle\,\pi(\vec q\,)\pi(-\vec
q\,)\,|\,J(0)\,|\,0\,\rangle \vert \cos \delta(2 E_q)$.  This can be
achieved by working in a finite volume in the following way.
Considerations similar to those in section~\ref {sec:large} give
\begin{equation}
{\cal G}_3(t, \vec q\,)= \sqrt{2E_qV}\sum_{n} A_V(E_n) e^{- (E_n-E_q)\,t}
\mathrel{\mathop{\kern0pt\longrightarrow}\limits_{V\to \infty }}
\int\limits_{2 \, m_\pi}^\infty \, dE \,  A_\infty(E)
e^{-(E- E_q)t} \,,
\label{eq:domenica}\end{equation}
where the finite-volume spectral function $A_V$ is defined as
\begin{equation}
A_V(E_n)\equiv\, _V\langle\,\pi(\vec q)\,|\,\Phi_{-\vec q\,}\,|
\pi\pi,n\rangle_V\ _V\langle\,\pi\pi,n\,|\,J(0)\,|\,0\,\rangle\ .
\label{eq:factors}\end{equation} 
Eq.~(\ref{eq:cosdelta}) allows us to write
\begin{eqnarray}
\int\limits_{2 \, m_\pi}^\infty \, dE \,  A_\infty(E)
e^{-(E- E_q)t} &=&\nonumber\\ 
&&\hspace{-2in}
\frac{\sqrt{Z}}{2E_q}\, e^{-E_qt}\
\big|\,_{\textrm{out}}\langle\,\pi(\vec q\,) \pi(-\vec
q\,)\,|\,J(0)\,|\,0\,\rangle\,\big|\,\cos \delta(2 E_q)
+
\int\limits_{2 \, m_\pi}^\infty \, dE \,  A^{P.V.}_\infty(E)
e^{-(E- E_q)t}\,,
\end{eqnarray}
where $P.V.$ denotes the contribution coming from the principal value
of the integral. Using the Poisson summation formula in eq.~(\ref{poisson}),
we obtain
\begin{eqnarray}
{\cal G}_3(t, \vec q\,)&
\mathrel{\mathop{\kern0pt\longrightarrow}\limits_{V\to \infty }}&
\frac{\sqrt{Z}}{2E_q}\, e^{-(E_{\tilde n}-E_q)t}\
\big|\,_{\textrm{out}}\langle\,\pi(\vec q\,) \pi(-\vec
q\,)\,|\,J(0)\,|\,0\,\rangle\,\big|\,\cos \delta(2 E_q)
\nonumber\\ 
&+&
\sum_n \frac{A^{P.V.}_\infty(E)}{\rho_V(E_n)}
e^{-(E_n-E_q)t}\, + O\left(\frac{1}{L}\right)\,,
\label{eq:martedi}\end{eqnarray}
where $E_{\tilde n}$ is the closest approximation to $2E_q$ among the
solutions of eq.~(\ref{phase}). The replacement of $2E_q$ by
$E_{\tilde n}$, at fixed $t$, is valid up to power corrections in the
volume. Due to the presence of $\rho_V$ in the denominator, each term
in the sum in eq.~(\ref{eq:martedi}) vanishes at large volumes (as
inverse powers of the volume). Thus, from a numerical evaluation of
${\cal G}$ we obtain
\begin{equation}
A_V(E_{\tilde n}) = \frac{1}{\sqrt{2E_qV}}\,\frac{\sqrt{Z}}{2E_q}\,
\big|\,_{\textrm{out}}\langle\,\pi(\vec q\,) \pi(-\vec
q\,)\,|\,J(0)\,|0\rangle\big|\,\cos \delta(2 E_q) \ ,
\label{eq:nuova}\end{equation}
up to power corrections in the volume. Eq.~(\ref{eq:nuova}) was
suggested in ref.~\cite{ciuchini}, under a smoothness assumption on
the amplitudes.

We end this section by explaining that the conclusion of the
preceeding paragraphs is consistent with the results of
ref.~\cite{lll} and the discussion in the previous sections. Consider
eq.~(\ref{eq:factors}). From the final-state interaction theorem, we
deduce that the phases of the two factors are equal and opposite so
that the right-hand side is equal to the product of the absolute
values. By an additional measurement of the four-pion correlation
function (see, for example, Appendix D), we can obtain
$|_V\langle\,\pi(\vec q)\,|\,\Phi_{-\vec q\,}\,|
\pi\pi,n\rangle_V|$, and hence determine $|_V\langle\,\pi\pi,n\,
|\,J(0)\,|\,0\,\rangle|$. Correcting this finite-volume matrix
element by the LL-factor we determine the infinite-volume matrix
element up to exponentially small corrections. This is not the
procedure which was followed in arriving at the real part of the
matrix element in eq.~(\ref{eq:nuova}) however, and this explains
the different precision of the two cases. 

\section{Conclusions}
\label{sec:conclusion}

In this paper we have discussed the conditions under which the
amplitudes of two-body decays below the inelastic threshold, such
as $K\to\pi\pi$ decays, can be computed in a finite volume.  We
have shown that the correction factor recently found by Lellouch
and L\"uscher~\cite{lll}, which relates the finite-volume matrix
elements and the physical (infinite-volume) decay amplitudes at
zero momentum transfer for finite volume states up to $n=7$, can
be extended, with exponential precision in the volume, to all
elastic states under the inelastic threshold and to any momentum
transfer.

We have examined the possible influence of inelastic thresholds both
on these correction factors and on the quantization condition for
two-particle states (below the inelastic threshold) in a finite
volume.  Our conclusion is that, in general, the presence of inelastic
thresholds requires us to work on sufficiently large volumes so that
the sum over elastic states under the inelastic threshold can be
approximated reliably by the corresponding integral over the
energy. In lattice simulations in the foreseeable future the
fulfillment of this condition will represent a formidable challenge.
It is therefore reassuring that in hadronic sectors, such as the
two-pion system in $K\to\pi\pi$ decays, where the interaction is
rather weak and inelasticity develops fully only at energies
significantly higher than the kinematic thresholds, the finite-volume
approach may be applicable starting from relatively small volumes. The
quantitative implications for lattice computations of $K\to\pi\pi$
decay amplitudes will need to be investigated in numerical
simulations.

We have also examined a different and frequently used method to obtain
the real part of the $K \to \pi
\pi$ amplitude based on the correlator $\langle 0\, |\,T[ \pi \pi
{\cal H}_{W} K]\,|\,0\rangle $. We find that in
this case there are finite-volume corrections, which vanish only as 
inverse powers of the volume, and we have explained the connection
with the approach of ref.~\cite{lll}. 

Nonleptonic weak decays will continue to play a central r\^ole in
particle physics phenomenology in the coming years.  We trust that
this paper is a contribution to the theoretical framework which will
underpin lattice simulations of $K\to\pi\pi$ decays.  Further work is
needed in order to understand how to treat contributions from states
above the inelastic threshold, particularly in relation to nonleptonic
two-body $B$-decays.  For this important class of processes a huge
amount of experimental data is becoming available and yet its
interpretation is currently limited by our inability to quantify the
strong interaction effects.

\subsection*{Acknowledgements} We warmly thank Laurent Lellouch and
Martin L\"uscher for many helpful discussions throughout the course of
this work. We gratefully acknowledge interesting discussions with our
colleagues from the EU network on Hadron Phenomenology, and especially
with Peter Hasenfratz and Ferenc Niedermeyer.

This work was supported by European Union grant HTRN-CT-2000-00145.
DL and CTS acknowledge support from PPARC through grants
PPA/G/S/1997/00191, PPA/G/O/\-1998/\-00525 and PPA/G/S/1998/00529.

\section*{Appendix A} In this appendix we show that
${\cal G}(t,\vec q)$, defined in eq.~(\ref{eq:gdef}), is equal to
the expression given in eq.~(\ref{eq:mtresult}) also in Minkowski
space. We will see that although the right-hand side of
eq.~(\ref{eq:mtresult}) contains the average of the matrix
elements into \textit{in} and \textit{out} two-pion states, the
term containing the principal value integral is precisely that
required to recover the full physical amplitude in Minkowski
space, but not in Euclidean space. The aim of this discussion is
to clarify the the similarities and differences between the
Euclidean and Minkowski correlation functions.

We start by inserting a complete set of \textit{out}-states
$\{|m\rangle \}$~\footnote{ As before we restrict our discussion
to two-pion intermediate states.} in ${\cal G}_3$, defined in
eq.~(\ref{eq:g3defb}), and obtain
\begin{equation}
{\cal G}_3(t,\vec q)=\sum_{m}(2\pi)^3\delta^{(3)}(\vec p_m)\,
e^{-i(E_m-E_q)t}\,\langle\pi(\vec q)\,|\,\Phi_{\meno\vec
q\,}(0)\,|m\rangle \langle m\,|\,J(0)\,|\,0\rangle\ ,
\label{eq:g3mink}\end{equation}
where $\vec p_m$ is the momentum of the two-pion state labeled by $m$.

The disconnected contribution to $G_3$ is readily obtained using
\begin{equation}
\langle\pi(\vec q\,)\,|\,\Phi_{\meno\vec q\,}(0)\,|\pi(\vec k\,)\,
\pi(\meno\vec k\,)\,\rangle_{\out}^{\disc}=2E_q\,(2\pi)^3\,
\delta^{(3)}(\vec q\,\meno\vec k\,)\,\sqrt{Z}\ ,
\label{eq:disc}\end{equation}
so that
\begin{equation}
{\cal G}^{\disc}_3(t,\vec q\,)=\frac{\sqrt{Z}}{2E_q}\,e^{-iE_qt}\,
_{\out}\langle\,\pi(\vec q\,)\pi(\meno\vec q\,)\,|\,J(0)\,|0\rangle\ .
\end{equation}

To evaluate the connected contribution we use the reduction formula
\begin{eqnarray}
_{\out}\langle m\,|\,\pi(q_1) \pi(q_2)\,\rangle_{\inn} &=&2E_2 \
_{\out}\langle m\,\meno \pi(q_2)\,|\,\pi(q_1)\,\rangle_{\inn}
\nonumber\\ && \hspace{-1.5in}+\frac{i}{\sqrt{Z}}\,(2\pi)^4
\delta^{(4)}(p_m\,\meno q_1\,\meno q_2) \,(-p^2+m_\pi^2)\
_{\out}\langle m\,|\phi (0)\,|\,q_1\,\rangle\ ,
\label{eq:reduction}\end{eqnarray} where temporarily we introduce
general momenta $q_1$ and $q_2$. $\langle m-\pi(q_2)|$ represents
the state containing the particles in state $\langle m|$ with
$\pi(q_2)$ removed.  This contribution is absent if $\langle m|$
does not contain such a pion.  The scattering amplitude ${\cal
M}(\vec q_1, \vec q_2;m)$ is defined from the limit as
$p=(E_m\meno E_1,\vec q_2\,)$ goes on-shell, $p^2\to m_\pi^2$,
\begin{equation}
\langle \pi(\vec q_1)\,|\phi(0)\,|\,m\rangle^{\conn}_{\out}
=\sqrt{Z}\frac{{\cal M}^\ast}{-p^2+m_\pi^2+i\varepsilon}=\meno\sqrt{Z}
\frac{{\cal M}^\ast}
{(p_0-E_2-i\varepsilon)(p_0+E_2+i\varepsilon)}\ .
\end{equation}
The physical scattering amplitude for $\pi(q_1)+\pi(q_2)\to m$ is
therefore given by
\begin{equation}
(2\pi)^4\delta^{(4)}(p_m-q_1-q_2)\,\big(i{\cal M}(\vec q_1,\vec q_2;m)\big)
=\ _{\out}\langle m\,|\pi(q_1)\pi(q_2)\,\rangle_{\inn}\ \meno\
_{\out}\langle m\,|\pi(q_1)\pi(q_2)\,\rangle_{\out}\ ,
\label{eq:m}\end{equation}
where we have used the fact that single-particle $in$ and $out$ states
are the same.

We now use
\begin{equation}
\frac{1}{p_0\,\meno E_2\meno i\varepsilon}=i\pi\,\delta(p_0\,\meno
E_2) + {\cal P}\frac{1}{p_0\,\meno E_2}\ ,
\end{equation}
where ${\cal P}$ stands for ``principal value", to write
\begin{equation}
\langle \pi(\vec q_1)\,|\phi(0)\,|\,m\rangle^{\conn}_{\out} =\,\meno\,
i\pi\,\frac{\sqrt{Z}}{2E_2}\,{\cal M}^\ast\,\meno\frac{\sqrt{Z}\,{\cal M}^\ast}
{p_0+E_2+i\varepsilon}\,{\cal P}\frac{1}{p_0-E_2}\ .
\label{eq:sep}\end{equation}

The first term on the r.h.s.  of (\ref{eq:sep}) gives a contribution
to ${\cal G}_3$ of
\begin{equation}
{\cal G}_3^{\conn ,A}= \frac{\sqrt{Z}}{2E_2}\, e^{-iE_2 t_2}
\times\frac12\Big[\ _{\inn}\langle\pi(q_1)\pi(q_2)\,|J(0)\,|0\rangle
\,\meno\ _{\out}\langle\pi(q_1)\pi(q_2)\,|J(0)\,|0\rangle\Big]\ ,
\end{equation}
which, when combined with the disconnected term gives
\begin{equation}
{\cal G}_3^{\disc}+{\cal G}_3^{\conn ,A}= \frac{\sqrt{Z}}{2E_2}\,
e^{-iE_2 t_2} \times\frac12\Big[\
_{\inn}\langle\pi(q_1)\pi(q_2)\,|J(0)\,|0\rangle \,+\
_{\out}\langle\pi(q_1)\pi(q_2)\,|J(0)\,|0\rangle\Big].
\end{equation}
Thus we see that \textit{even in Minkowski space} we get the average
of the matrix elements into $in$ and $out$-states.  The remaining
contribution to ${\cal G}_3$ comes from the principal value term in
(\ref{eq:sep}).

In Minkowski space however, writing the result in the form of
(\ref{eq:mtresult}) is not very transparent. To recover the standard
result recall that ${\cal G}_3$ contains a factor of $\exp(-iE_m t)$
and that
\begin{equation}
\langle\pi(\vec q\,)\,|\Phi_{\meno\vec q\,}(0)\,|m\rangle_{\out}^{\conn}
=-\frac{\sqrt{Z}{\cal M}^\ast}{p^2-m_\pi^2-i\varepsilon}=
-\frac{\sqrt{Z}{\cal M}^\ast}{E_m(E_m-2E_q-i\varepsilon)}\ .
\end{equation}

\par The sum over the intermediate states can be written as an
integral over $E_m$.  As $t \to \infty$ this integral, which is
sufficiently convergent, can be extended to the range
$(-\infty,\infty)$ and evaluated by contour integration.  For positive
$t$ we can close the contour in the lower half-plane and, since there
are no singularities there, we obtain zero for the connected
contribution.  In this way we obtain the standard result
\begin{equation}
{\cal G}^{\textrm{Minkowski}}(t_1,t_2;\vec q)=\frac{Z}{(2E_q)^2}\,
e^{-iE_q(t_1+t_2)}\ _{\out}\langle\pi(\vec q)\pi(\meno\vec
q)\,|\,J(0)\,|0\rangle\ .
\end{equation}

\par In Euclidean space it is not possible to treat the connected part
in this way.  Instead of the factor of $\exp(-iE_m t)$, ${\cal G}_3$
now has one of $\exp(-E_m t)$ which, at large times $t$ is dominated
by $E_m\simeq 2m_\pi$ i.e. the lowest energy state consistent with
three-momentum conservation.  It is therefore not possible to perform
the energy integration by contours.

\section*{Appendix B} In this appendix we present a derivation of
eq.~(\ref{eq:cosdelta}) based on elastic unitarity. The starting point
once again is the quantity
\begin{equation}
{\cal G}_3(t,\vec q)\equiv\langle\pi(\vec q\,)\,| \Phi_{-\vec
q\,}(t)\,J(0)\,|\,0\rangle \, , \label{eq:g3defc}\end{equation}
defined in eq.~(\ref{eq:g3defb}) of section~\ref{sec:delta}. From
the discussion in  sec.~\ref{sec:mt} and appendix A we know that
${\cal G}_3$ is given by:
\begin{eqnarray}
{\cal G}_3(t,\vec q\,) & = & \frac{\sqrt{Z}}{2E_q}\, e^{-E_qt}\
_{\textrm{out}}\langle\,\pi(\vec q\,)\pi(-\vec
q\,)\,|\,J(0)\,|0\rangle\times \Big[\,1-\frac{i}{2}{\cal M}^\ast(2
E_q)\,{\cal N}(2 E_q)\,\Big] +\nonumber\\
&&\hspace{-1.2in}{\cal P}\,\int\frac{dE}{2\pi}\,{\cal N}(E)\,
\langle\,\pi(\vec q\,)\,|\,\Phi_{-\vec q\,}(0)\,|\pi(\vec
k\,)\,\pi(-\vec k)\,\rangle_{\textrm{out}} ^{\textrm{conn}}\
_{\textrm{out}}\langle\,\pi(\vec k\,)\pi(-\vec
k\,)\,|\,J(0)\,|0\rangle\, e^{-(E-E_q)t}\,,
\label{eq:g3interp}\end{eqnarray}
where the phase space factor ${\cal N}(E)$ is given by
\begin{equation}
{\cal N}(E)=\frac{k(E)}{4\pi E}\, ,
\end{equation}
and the scattering amplitude ${\cal M}(2 E_q)\equiv{\cal M}(\vec
q,-\vec q;\vec q^{\,\prime},-\vec q^{\,\prime})$ ($\vert q^\prime
\vert = \vert q\vert$) is given by eq.~(\ref{eq:m})
\begin{equation}
(2\pi)^4\delta^{(4)}(p_m-q_1-q_2)\,\big(i\, {\cal M}(\vec q_1,\vec
q_2;m)\big) =\ _{\out}\langle
m\,|\pi(q_1)\pi(q_2)\,\rangle_{\inn}\ \meno\ _{\out}\langle
m\,|\pi(q_1)\pi(q_2)\,\rangle_{\out}.
\end{equation}
Since $J$ creates a two-pion state of fixed isospin  and angular
momentum from the vacuum, we have
\begin{eqnarray}
_{\textrm{out}}\langle\,\pi(\vec q\,)\pi(-\vec
q\,)\,|\,J(0)\,|\,0\,\rangle = e^{i\delta(2 E_q)}\
\big|\,_{\textrm{out}}\langle\,\pi(\vec q\,)\pi(-\vec
q\,)\,|\,J(0)\,|\,0\,\rangle\, \big| \, .  \label{eq:60}
\end{eqnarray}
On the other hand, the elastic scattering amplitude is given, in terms
of the scattering phase, by
\begin{equation}
{\cal M}(2 E_q)=\frac{2}{{\cal N}(2 E_q)}\,e^{i\delta(2 E_q)}\,
\sin(\delta(2 E_q))\ ,
\end{equation}
and satisfies the unitarity relation
\begin{equation}
2\,\textrm{Im}[{\cal M}(2 E_q)]={\cal N}(2 E_q)\,\big|{\cal M}(2 E_q)\big|^2\
.
\label{eq:62} \end{equation}
\par Using eqs.~(\ref{eq:60})--(\ref{eq:62}), we obtain
\begin{equation}
{\cal G}_3(t, \vec q\,)= \frac{\sqrt{Z}}{2E_q}\, e^{-E_qt}\
\big|\,_{\textrm{out}}\langle\,\pi(\vec q\,) \pi(-\vec
q\,)\,|\,J(0)\,|0\rangle\,\big|\,\cos \delta(2 E_q)\,+\cdots\ ,
\label{eq:cosdeltab}\end{equation}
where the ellipses represent the principal value term in
eq.~(\ref{eq:g3interp}).

\section*{Appendix C}

In this appendix we reproduce some of the results of
ref.~\cite{ml} in a form suitable for our presentation.  We recall
that we are studying the eigenvalue problem for a two particle
system of mass $m$ in the center of mass, subject to a repulsive
spherically symmetric potential $V(r)$.  For simplicity, we assume
that $V(r)$ only gives rise to an s-wave scattering phase shift
$\delta(k)$\footnote{For the treatment of the most general case we
refer the reader to ref.~\cite{ml}.}, where $k$ denotes the center
of mass relative momentum of the two particle system.  We also
assume that the potential has a finite range $R$, outside of which
it vanishes identically.  Our problem is to find the structure of
the eigenvalues $k^{2}$ and the corresponding eigenfunctions of
the Schroedinger equation
\begin{equation}
(\Delta +k^{2}) \psi(\vec x) = m V(r) \psi(\vec x)
\label{schr}
\end{equation}
with periodic boundary conditions in a cubic box with sides of
length $L>R$. In particular we are interested in the cubically
invariant eigenfunctions with non-zero s-wave projections; these
are the only ones which can contribute to the sum over
intermediate states in eq.~(\ref{equal}).

As a result of the hypothesis on the range of the potential, for
$R \leq r \leq \frac{L}{2}$, eq.~(\ref{schr}) reduces to the
Helmholtz equation
\begin{equation}
(\Delta +k^{2}) \psi(\vec x) =0\ . \label{Helmholtz}
\end{equation}
We start by considering the case $k^{2} \neq \frac {4
\pi^{2}}{L^{2}} \vec n^{2}$, where $\vec n$ denotes any vector
with integer components.  In this case there is no solution which
is valid throughout the whole volume.

In order to treat this problem we need the finite volume Green
function
\begin{equation}
(\Delta +k^{2}) G_{V}(\vec x-\vec x')=\delta (\vec x-\vec x')\ ,
\end{equation}
given by
\begin{equation}
G_{V}(\vec x-\vec x')=\frac{1}{V} \sum_{\{ \vec p_{n}\}}\frac{e^{i
\vec p_{n} \cdot(\vec x-\vec x')}}{k^2-\vec p_{n}^{2}}\ ,
\label{eq:fvgf}\end{equation}
where
\begin{equation}
\vec p_{n}=\frac {2 \pi}{L} \vec n\ .
\end{equation}
It is convenient to consider the expansion of $G_{V}(\vec x)$
in spherical harmonics.  In particular we will be interested in the
spherical projection of $G_{V}(\vec x)$
\begin{equation}
G_{V}^{(0)}(r)\equiv \frac{1}{4 \pi} \int_{\Omega} d\Omega \,\,
G_{V}(\vec x)=\frac{1}{V} \sum_{\{ \vec p_{n}\}}\frac{1}{k^{2}-
p_{n}^{2}}\frac{\sin p_{n} r}{p_{n} r}
\end{equation}
with $p_{n}\equiv |\vec p_{n}|$.  For $r\neq 0$, $G_{V}^{(0)}(r)$
satisfies the Helmholtz equation, so that we have
\begin{equation}
G_{V}^{(0)}(r)=- \frac {\cos kr}{4 \pi r}+c \, \,\frac {\sin kr}{k
r}\ . \label{spherical}
\end{equation}
The term $- \frac {\cos kr}{4 \pi r}$ in eq.~(\ref{spherical}) by
itself satisfies:
\begin{equation}
(\Delta +k^{2}) (- \frac {\cos kr}{4 \pi r}) =\delta
(\vec x-\vec x')
\end{equation}
so that the second term on the r.h.s. of eq.~(\ref{spherical})
must be regular. A possible way to determine $c$ is from the
relation
\begin{equation}
c=\lim_{r\to 0} \frac {d}{dr} (r G_{V}^{(0)}(r))
\end{equation}
so that eq.~(\ref{spherical}) gives
\begin{equation}
c=\lim _{r \to 0} \frac{1}{V} \sum_{\{ \vec p_{n}\}}\frac{\cos
p_{n} r}{k^{2}-\vec p_{n}^{2}}\ .\label{correction1}
\end{equation}
We add a note of caution at this point. The sum $\frac{1}{V}
\sum_{\{ \vec p_{n}\}}\frac{1}{k^{2}-\vec p_{n}^{2}}$ is formally
the same as $G_{V}(0)$ and is therefore divergent. The limit in
eq.~(\ref{correction1}) could provide the necessary procedure for
the analytic continuation (analogous to $s$-analytic continuation
in ref.~\cite{ml}). We will not expand further on
this point and we note that eq.~(\ref{spherical}), together with
the value of $c$ given in eq.~(\ref{correction}) below, coincides
with eq.(3.29) in the second paper of ref.\cite{ml}.

Eq.~(\ref{correction1}) together with the definition
\begin{equation}
Z_{00}(s,q^{2}) \equiv \frac {1}{\sqrt {4 \pi}} \sum_{\{ \vec
n\}}\frac{1}{(\vec n^{2} - q^{2})^{s}}
\label{eq:z00def}\end{equation}
gives
\begin{equation}
c= - \frac {Z_{00}(1,q^{2})}{2 \pi ^{\frac{3}{2}} L}\ .
\label{correction}\end{equation}

When $R\leq r \leq \frac {L}{2}$, the potential is zero and the
wave function propagates according to the Helmholtz equation
eq.~(\ref{Helmholtz}).  Then, as usual, knowledge of the Green
function $G_{V}(\vec x-\vec x')$ implies that
\begin{equation}
\psi(\vec x ')=\int_{V-S_{R}} d^{3} x \, \, \vec
\partial_{x} \cdot [\psi(\vec x) \vec \partial_{x}
G_{V}(\vec x -\vec x ')- \vec \partial_{x}
\psi(\vec x) G_{V}(\vec x-\vec x ')]
\end{equation}
where the integral is extended to the cubic volume with the sphere
of radius $R$ subtracted. Using Gauss' theorem and the periodicity
in $V$, we have
\begin{equation}
\psi(\vec x ')= R^{2} \int_{S_{R}} d \sigma \, \,
[\psi(\vec x) \vec \partial_{x}
G_{V}(\vec x-\vec x ')- \vec \partial_{x}
\psi(\vec x) G_{V}(\vec x -\vec x ')] \cdot
\vec n
\label{propagation}
\end{equation}
where $\vec n$ is the internal normal to the sphere $S_{R}$.
$\psi(\vec x)$ itself is a very complicated object, the
superposition of an infinite number contributions from states of
different angular momenta.
The quantization condition is that its s-wave projection must
coincide with the unperturbed s-wave wave function in infinite
volume:
\begin{equation}
f(r)= A \frac {\sin k r} {k r} -B \frac {\cos k r} {k r}
\end{equation}
where $A$ and $B$ are related to the s-wave phase shift $\delta$
by
\begin{equation}
\tan \delta =\frac {B} {A}\ .
\end{equation}
Here we have used the condition that only the s-wave component is
interacting. In general, the $l$-th partial wave contributes a
radial function:
\begin{equation}
f_{l}(r)=A_{l} \, j_{l}(kr) + B_{l} \, n_{l}(kr)
\label{radial}
\end{equation}
where $j_{l}(kr)$ is regular everywhere, while $n_{l}(kr)$ is
singular at the origin and signals the presence of the interaction
\begin{equation}
\tan \delta_{l} =\frac {B_{l}} {A_{l}}\ .
\end{equation}

Since we assume that only the s-wave component is interacting, it
is only for the s-wave that $B \neq~0$. Now
if we set $B_{l}=0$ on the r.h.s.  of eq.~(\ref{propagation}), the
result must be exactly zero, otherwise we would find regular
solutions of the Helmholtz equation with correct boundary
conditions, which is impossible if $k^{2} \neq \vec p_{n}^{2}$ for
all $\vec p_{n}$.  This means that in this case we can put to zero
all contributions from states with $l \neq 0$ on the r.h.s. of
eq.~(\ref{propagation}).

Since the only contribution to the r.h.s. of
eq.~(\ref{propagation}) comes from the rotationally invariant
component of $\psi(\vec x)$ and we are interested in the
projection of eq.~(\ref{propagation}) onto the s-wave, we require
the angular average of the Green function $G_{V}^{(0)}(| \vec x -
\vec x ' |)$ with respect to $\vec x$ or $\vec x '$:
\begin{equation}
G^{(0,0)}_{V}(r,r') \equiv \int_{\Omega '} \frac {d \Omega}{4 \pi}
G^{(0)}_{V}\left(\sqrt {r^{2}+{r'}^{2} - 2 r r' \cos \theta}\,\right).
\end{equation}
Evaluation of the integral gives
\begin{equation}
G^{(0,0)}_{V}(r,r') = \frac {\sin k |r-r'|- \sin k (r+r')}{8 \pi k
r r'} - c \frac {\cos k (r+r') - \cos k (r-r')}{2 k^{2} r r'}\ .
\end{equation}
The s-wave projection of eq.~(\ref{propagation}) then gives rise to
the quantization condition
\begin{equation}
f(r)=-4 \pi R^{2} [f(R) \frac {d}{d R} G^{(0,0)}_{V}(R,r') -f'(R)
G^{(0,0)}_{V}(R,r')]\hspace{0.2in}\textrm{with}\hspace{0.2in}r'\geq
R\ .\label{eigen}
\end{equation}
After some elementary algebraic manipulations eq.~(\ref{eigen})
reduces to
\begin{equation}
A \frac {\sin k r} {k r} -B \frac {\cos k r} {k r}= -\frac {B
(k\cos kr - 4 \pi c \sin k r)} {k^{2} r}\ ,
\end{equation}
from which the L\"uscher quantization condition, eq.~(\ref{phase})
follows immediately.

The derivation sketched here shows that the L\"uscher quantization
formula enumerates all the states with $k^{2} \neq \frac {4
\pi^{2}}{L^{2}} \vec n^{2}$ that have a non-zero projection on the
s-wave.  These states are non-degenerate, since the final
eigenvalue equation has only one solution. There are other
eigenstates, not described by eq.~(\ref{phase}), which must be
considered; these include those with $k^{2} = \frac {4
\pi^{2}}{L^{2}} \vec n^{2}$ for some $ \vec n$. These states have
an energy which is independent of the interaction because we have
assumed that only the s-wave phase shift is non-zero~\footnote{In
the presence of fully interacting higher partial waves, these
states would have a non-trivial quantization condition which could
be found by the methods of ref.~\cite{ml} or by a natural
extension of the method presented in this appendix.}. They are
constructed as follows. We start with a plane wave of the form
$e^{i\vec p_{n} \cdot \vec x}$. This is not a solution of
eq.~(\ref{schr}) because it has an s-wave projection, so that the
potential $V(r)$ acts on it in non-trivially.  However the
combination
\begin{equation}
{\tilde\psi}(\vec x) = e^{i\vec p_{n} \cdot \vec x} - e^{i\vec
p^{\,\prime}_{n} \cdot \vec x}\,, \label{spur}
\end{equation}
where $\vec p_{n} \neq \vec p^{\,\prime}_{n}$, but
$p_{n} = p'_{n}$
is an exact solution of eq.~(\ref{schr}) with $k^{2} = \frac {4
\pi^{2}}{L^{2}} \vec n^{2}$, because it has zero projection on the
s-wave. Moreover if $\vec p^{\,\prime}_{n}$ is different from the
cubic rotations of $\vec p_{n}$, which is possible when $| \vec n
| \geq 8$, the state in eq.~(\ref{spur}) can be averaged to get a
cubically invariant eigenstate.
However, it is clear from this derivation that these states do not
contain an s-wave component and do not therefore contribute to the
sum over intermediate states in eq.~(\ref{equal}).

There is the further possibility of states corresponding to the
eigenvalue $k^{2} = \frac {4 \pi^{2}}{L^{2}} \vec n^{2}$
\underline{and} with a non-zero s-wave projection~\cite{ml}. The
condition for the presence of these states can be found using the
procedure outlined before. Since the eigenvalue is fixed a priori,
the corresponding eigenvalue equation represents a consistency
condition in the form of a relation between $\vec n^{2}$, the
phase shifts and the volume $V$. These states therefore are not
generic and can only occur for particular volumes and/or
potentials and we therefore do not include them in our argument.

We end this appendix by briefly showing that the knowledge of the
finite-volume Green function eq.~(\ref{eq:fvgf}) and its projection on
a given partial wave (such as the s-wave projection in
eq.~(\ref{spherical})\,) allows one to readily reproduce the class of
summation formulae given in eq.(2.51) of the second paper of
ref.~\cite{ml}. These formulae are central in computing finite-volume
corrections in perturbation theory in quantum mechanics and field
theory. For illustration we will discuss the particular case of
a spherically symmetrical function $\tilde f(p^2)$, for which we 
prove below that:
\begin{equation}
\frac{1}{V}{\sum_{\{\vec p_n\}}}\frac{\tilde f(p_n^2)}
{k^2-p_n^2}=\frac{1}{(2\pi)^3}{\cal P}\int_{-\infty}^{\infty}d^{\,3}p\,
\frac{\tilde f(p^2)}{k^2-p^2}\,+\,c\tilde f(k^2)
\label{eq:summation}\end{equation}
up to exponentially small
corrections, provided that $f(r)$, the Fourier tranform of $\tilde f(p^2)$,
\begin{equation}
f(r)\equiv\frac{1}{(2\pi)^3}\int_{-\infty}^{\infty}d^{\,3}p\,
e^{-i\vec p\cdot\vec x}\tilde f(p^2)\ ,
\end{equation}
vanishes exponentially with $r\equiv|\vec x|$. In
eq.~(\ref{eq:summation}) $p_n\equiv|\vec p_n|$ and $k^2$ is different
from any of the allowed $p_n^2$ on the given volume. $c$ is defined in
eq.~(\ref{correction}) as a function of $k^2$ and $L$. To prove
eq.~(\ref{eq:summation}) we start with the finite-volume integral:
\begin{eqnarray}
\int_Vd^{\,3}xf(r)G_V(\vec x)&=&\int_Vd^{\,3}xf(r)G_V^{(0)}(r)\nonumber\\ 
&=&\int_Vd^{\,3}xf(r)\left(-\frac{\cos(kr)}{4\pi r}+c\frac{\sin(kr)}{kr}
\right)\ .
\label{eq:summation1}\end{eqnarray}
The assumption of the exponential decrease of $f(r)$ at large $r$ allows
us to extend the integrals in eq.~(\ref{eq:summation1}) to infinite volume
with an exponentially small error in the volume, so that:
\begin{eqnarray}
\int d^{\,3}xf(r)G_V(\vec x)&=&
\frac{1}{V}{\sum_{\{\vec p_n\}}}\frac{\tilde f(p_n^2)}{k^2-p_n^2}\,,\\ 
\int d^{\,3}xf(r) \left(-\frac{\cos(kr)}{4\pi r}\right)&=&
\frac{1}{(2\pi)^3}{\cal P}\int_{-\infty}^{\infty}d^{\,3}p\,
\frac{\tilde f(p^2)}{k^2-p^2}\,,\\ 
\int d^{\,3}xf(r)\frac{\sin(kr)}{kr}&=&\tilde f(k^2)\,.
\end{eqnarray}
This proves eq.~(\ref{eq:summation}).

Eq.~(\ref{eq:summation}) can also be used in the limit as $k^2$
approaches one of the $p_n^2$, $\vec K^2$ say. In this limit both
sides of eq.~(\ref{eq:summation}) become singular and we can write:
\begin{equation}
\frac{1}{V}{\sum_{\{\vec p_n\}}}^\prime\frac{\tilde f(p_n^2)}
{k^2-p_n^2}=\frac{1}{(2\pi)^3}{\cal P}\int_{-\infty}^{\infty}d^{\,3}p\,
\frac{\tilde f(p^2)}{k^2-p^2}\,+\lim_{k\to|\vec K|} 
\left\{c\tilde f(k^2)-\frac{\nu_K}{V}\frac{\tilde f(K^2)}{k^2-K^2}\right\}\ ,
\label{eq:singsummation}\end{equation}
where, as usual, $\nu_K$ is the number of $\vec p_n$'s with $p_n^2=K^2$
and the prime on the summation indicates the omission of the $\nu_K$ terms 
with $p_n^2=K^2$. From eq.~(\ref{correction}) we have that 
\begin{equation}
c=\frac{\nu_K}{V}\frac{1}{k^2-K^2}+c^{\textrm{reg}}\,,
\end{equation}
where $c^{\textrm{reg}}$ is defined as in eq.~(\ref{correction})
but with the $\nu_K$ terms with $n^2=n_K^2\equiv(LK/2\pi)^2$ excluded from the
sum in eq.~(\ref{eq:z00def}). $c^{\textrm{reg}}$ is therefore
regular in the limit $k^2\to K^2$,
\begin{equation}
\lim_{k^2\to K^2}c^{\textrm{reg}}=-\frac{z_K}{4\pi^2 L}\ ,
\end{equation}
where the number $z_K$ is defined by
\begin{equation}
z_K\equiv\sum_{\vec n^2\neq n_K^2}\frac{1}{\vec n^2-n_K^2}\ .
\end{equation}
Eq.~(\ref{eq:singsummation}) can now be rewritten:
\begin{equation}
\frac{1}{V}{\sum_{\{\vec p_n\}}}^\prime\frac{\tilde f(p_n^2)}
{K^2-p_n^2}=\frac{1}{(2\pi)^3}{\cal P}\int_{-\infty}^{\infty}d^{\,3}p\,
\frac{\tilde f(p^2)}{K^2-p^2}-\frac{z_K}{4\pi^2L}\tilde f(K^2)
+\frac{\nu_K}{V}\,\tilde f^\prime(K^2)\ ,
\label{eq:singsummation1}\end{equation}
where $\tilde f^\prime(k^2)$ represents the derivative of $\tilde
f(k^2)$ with respect to $k^2$.  Eq.~(\ref{eq:summation}) and
(\ref{eq:singsummation1}) can be easily extended to functions with an
angular dependence and to higher order poles. We stress also that if
we take $V\to\infty$ keeping the physical momentum $\vec K$ fixed (by
appropriately tuning the sequence of volumes to make this possible)
then we must consider $n_K$, $z_K$ and $\nu_K$ to be functions of the
volume and therefore the finite-volume corrections are not simply
given by the explicit factors of $1/L$ and $1/V$ on the right-hand
side of eq.~(\ref{eq:singsummation1}).

\section*{Appendix D}
In this paper we have seen examples of correlation functions from
which one can extract the modulus of the amplitude and others from
which one can obtain the modulus of the amplitude times
$\cos(\delta(2E_q))$.  The physical information which can be
obtained depends on the correlation function which has been
computed.  In this final appendix we present one more example,
demonstrating that $\cos (\delta(2 E_q))$ can be determined
directly from the evaluation of four-point correlation functions.

\par We start with the
following correlation function
\begin{equation}
{\cal G}_4(t_1,t_2,t_3,t_4)\equiv\langle\,0\,|\, \Phi_{\vec
q\,}(t_1)\,\Phi_{-\vec q\,}(t_2)\, \phi^\dagger(\vec
0,t_3)\,\phi^\dagger(\vec 0,t_4)\,|\,0\rangle\ ,
\label{eq:g4def}\end{equation}
with $t_1>t_2>t_3>t_4$.  As before we assume that we can neglect
inelastic contributions and that the flavour quantum numbers of the
interpolating operators are chosen so as to create two-pion states
with a fixed isospin.  For large $t_1$ and large negative $t_4$,
single-pion states dominate in the time intervals $(t_1,t_2)$ and
$(t_3,t_4)$ so that
\begin{equation}
{\cal G}_4(t_1,t_2,t_3,t_4)=\frac{\sqrt{Z}}{2E_q}\,e^{-E_qt_1}
\,\int\,\frac{d^{\,3}k}{(2\pi)^3}\ \frac{\sqrt{Z}}{2E_k}\, e^{-E_k|t_4|}
\,\langle\,\pi(\vec q\,)\,|\,\Phi_{-\vec q\,}(t_2)\,\phi(\vec
0,t_3)\,|\,\pi(\vec k)\, \rangle\ .
\end{equation}
Thus we need to consider the matrix element $\langle\,\pi(\vec
q\,)\,|\,\Phi_{-\vec q\,}(t_2)\,\phi(\vec 0,t_3)\,|\,\pi(\vec k)\,
\rangle$ to which we apply a similar  procedure to that in appendices A
and B. Inserting a complete set of (two-pion) states between the two
operators and applying the reduction formulae we find that
\begin{eqnarray}
{\cal G}_4(t_1,t_2,t_3,t_4)&=&\frac{Z^2}{4E_q^2}\,
e^{-E_q(t_1+t_2-t_3-t_4)}\,\big[1+\frac{i}{2}{\cal N}(2E_q){\cal M}(2E_q)\big]
\big[1-\frac{i}{2}{\cal N}(2E_q){\cal M}^\ast(2E_q)\big] \nonumber\\
&&\hspace{1in}\ +\ \textrm{P.V.}\nonumber\\
&=&\frac{Z^2}{4E_q^2}\,e^{-E_q(t_1+t_2-t_3-t_4)}\,
\cos^2(\delta(2E_q))\,+\,\textrm{P.V.}\ ,
\end{eqnarray}
where P.V. represents the principle value integral over energies other
than $2E_q$.  Thus from the evaluation of the four-pion correlation
function and the determination of the coefficient of the exponential
with exponent proportional to $E_q$ we may obtain the cosine of the
phase shift directly.

\end{document}